\documentclass[12pt]{article}
\usepackage{graphicx,epsfig,here}    
\begin{document}
\newcommand{\vsp}{\vspace*{-0.2cm}}
\thispagestyle{empty}

\begin{figure}[h]
\hspace{1.5cm}\includegraphics[width=4cm]{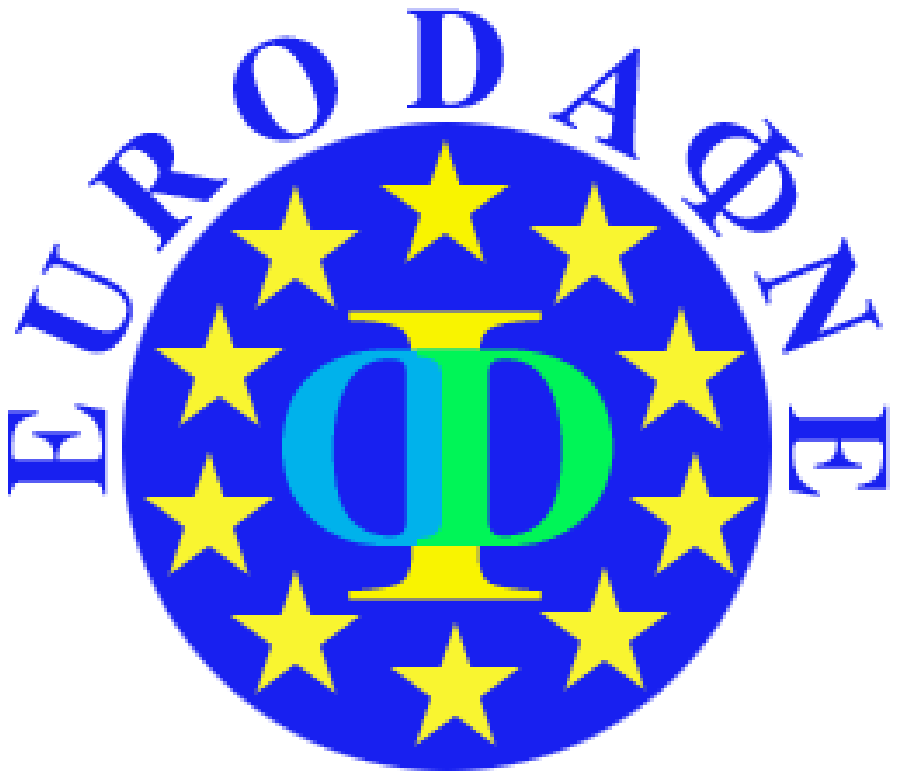}
\end{figure}

\vskip-2.5cm
\hspace{12cm}
\begin{tabular}{l}
BUTP-2001/23\\
BUHE-2001-07
\end{tabular}

\vskip 4cm

\begin{center}

{\Huge \bf HadAtom01}

\vskip 7mm

Workshop on Hadronic Atoms\\
Institut f\"ur Theoretische Physik, Universit\"at Bern\\
Sidlerstrasse 5, CH-3012 Bern, Schweiz\\
October 11 - 12, 2001\\
\end{center}
\vspace*{3mm}

\vskip 8mm
\begin{center}
{ J\"urg Gasser}\footnote{E-mail: gasser@itp.unibe.ch} \\
   {\it Institut f\"ur Theoretische  Physik, Universit\"at Bern,  
   CH-3012 Bern, Schweiz}
 \vskip 5mm
{ Akaki Rusetsky}\footnote{E-mail: rusetsky@itp.unibe.ch} \\
   {\it Institut f\"ur Theoretische  Physik, Universit\"at Bern,  
   CH-3012 Bern, Schweiz}\\
   {\it HEPI, Tbilisi State University, 380036 Tbilisi, Georgia}
 \vskip 5mm
{ J\"urg Schacher}\footnote{E-mail: schacher@lhep.unibe.ch} \\
   {\it Laboratorium f\"ur Hochenergiephysik, Universit\"at Bern,
   CH-3012 Bern, Schweiz}
\end{center}

\vskip 10mm
\begin{abstract}
\baselineskip 1.5em
These are the proceedings of the workshop "HadAtom01", held at
the Institut f\"ur Theoretische Physik, Universit\"at Bern,
October 11 - 12, 2001.
The main topics discussed at the  workshop were the physics of 
hadronic atoms and in 
this context recent results in experiment and theory.
Included here are the list of participants, the scientific program and
a short contribution from each speaker.
\end{abstract}


\newpage


\section{Introduction}


The workshop ``HadAtom01'' took place at the Institut f\"ur 
Theoretische
Physik, Universit\"at Bern on October 11-12, 2001. 
It was the third in a series of workshops on bound states, in particular
hadronic atoms: the previous
ones were held in Dubna, Russia (May 1998) [1], and in Bern (October
1999) [2].
The meeting was attended by about 50 physicists, and
results were presented by 25 participants. 

The topics of the workshop included

\vspace*{2mm}

\noindent
\begin{tabular}{l l l}
$\bullet$ Hadronic atoms, in particular their &~& $\bullet$ Experiments \\ 
\hspace*{4mm} Production &~& \hspace*{4mm} DIRAC at CERN \\ 
\hspace*{4mm} Interaction with matter &~~& \hspace*{4mm} DEAR at DAFNE \\ 
\hspace*{4mm} Energy levels &~& 
\hspace*{4mm} PSI (Pionic Hydrogen Collaboration) \\
\hspace*{4mm} Decays &~& \hspace*{4mm} Others \\ 
$\bullet$ Meson-meson and meson-baryon scattering &~& 
$\bullet$ $K_{\ell 4}$ decays \\
\end{tabular}

\vspace*{2mm}

The talks were devoted to recent experimental and theoretical progress in the
investigations of hadronic atoms. Among the highlights of the workshop were
the presentation of pre\-limi\-nary results of the DIRAC 
collaboration on the
measurement of the lifetime of pionium,
as well as the first measurement of kaonic nitrogen by the DEAR
collaboration. 

The speakers have provided a one-page abstract of the presentations 
as well as  a list of the most relevant references -
 these contributions are collected below.
We also display a list of the participants with their e-mail,
and the scientific program of the workshop.

\vskip.5cm

{ {\bf Acknowledgement} }

 We would like to thank 
all participants for their effort to
 travel to Bern and  
for making ``HadAtmo01''  an exciting and lively meeting.
We furthermore thank our secretaries,   Ruth
 Bestgen and Ottilia H\"anni,  for
 their excellent support,
Manuel Walker for designing the poster of the workshop,
and Christoph Haefeli, Andrea Macrina, Bernhard Scheuner, Julia Schweizer
  and Peter Zemp for their help during the workshop.
 Last but not least, we
 thank our colleagues from the organizing committee
 (Valery Lyubovitskij, 
 Leonid Nemenov, 
 Hagop Sazdjian, and  
 Dirk Trautmann)
 for their invaluable contribution in structuring the meeting.
This work was supported in part by the Swiss National Science
Foundation, and by TMR, BBW-Contract No. 97.0131  and  EC-Contract
No. ERBFMRX-CT980169 (EURODA$\Phi$NE).

\vspace*{.7cm}

 Bern, December 2001

\vskip.5cm

J\"urg Gasser, Akaki Rusetsky and J\"urg Schacher


\bigskip

\noindent\hrulefill

{\small

\begin{itemize}

\item[{[1]}]
Proceedings of the International Workshop ``Hadronic Atoms and Positronium in
the Standard Model'', Dubna, 26-31 May 1998, ed. M.A. Ivanov, A.B. Arbuzov,
E.A. Kuraev, V.E. Lyubovitskij, A.G. Rusetsky.
 
\vsp

\item[{[2]}]
J.~Gasser, A.~Rusetsky and J.~Schacher,
arXiv:hep-ph/9911339.
\end{itemize}

}


\newpage


\section{List of participants}

\vspace*{.3cm}

{\sf

\begin{center}

\renewcommand{\arraystretch}{0.85}

\begin{tabular}{r l l}
1. &Afanasev Leonid (Dubna)&leonid.afanasev@cern.ch\\
2. &Aste Andreas (Basel)&andreas.aste@unibas.ch\\
3. &Brekhovskikh Valeri (Protvino)&valeri.brekhovskikh@cern.ch\\
4. &B\"{u}ttiker Paul (J\"{u}lich)&p.buettiker@fz-juelich.de\\
5. &Colangelo Gilberto (Z\"{u}rich)&gilberto@physik.unizh.ch\\
6. &Cribeiro Alberto (Santiago)&alberto.cribeiro@rai.usc.es\\
7. &Cuplov Vesna (Marseille)&cuplov@cpt.univ-mrs.fr\\
8. &Curceanu [Petrascu] Catalina (Frascati)&catalina@lnf.infn.it\\
9. &Drijard Daniel (CERN)&daniel.drijard@cern.ch\\
10.&Egger Jean-Pierre (Neuch\^{a}tel)&jean-pierre.egger@unine.ch\\
11.&Eiras Dolors (Barcelona)&dolors@ecm.ub.es\\
12.&Gasser Juerg (Bern)&gasser@itp.unibe.ch\\
13.&Gianotti Paola (Frascati)&paola.gianotti@lnf.infn.it\\
14.&Goldin Daniel (Basel)&daniel.goldin@cern.ch\\
15.&Gotta Detlev (J\"{u}lich)&d.gotta@fz-juelich.de\\
16.&Guaraldo Carlo (Frascati)&guaraldo@lnf.infn.it\\
17.&Hasenfratz Peter (Bern)&hasenfra@itp.unibe.ch\\
18.&Heim Thomas (Basel)&thomas.heim@unibas.ch\\
19.&Hencken Kai (Basel)&k.hencken@unibas.ch\\
20.&Hennebach Maik (K\"{o}ln)&m.hennebach@fz-juelich.de\\
21.&Jensen Thomas (PSI)&thomas.jensen@psi.ch\\
22.&Knecht Marc (Marseille)&knecht@cpt.univ-mrs.fr\\
23.&Kubis Bastian (J\"{u}lich)&b.kubis@fz-juelich.de\\
24.&Lanaro Armando (CERN)&armando.lanaro@cern.ch\\
25.&Lipartia Edisher (Lund, Dubna \& Tbilisi)&lipartia@thep.lu.se\\
26.&Ludhova Livia (Fribourg)&livia.ludhova@unifr.ch\\
27.&Lyubovitskij Valery (T\"{u}bingen \& Tomsk)&
valeri.lyubovitskij@uni-tuebingen.de\\
28.&Mei\ss ner Ulf-G. (J\"{u}lich)&u.meissner@fz-juelich.de\\
29.&Nehme Abbass (Marseille)&nehme@cpt.univ-mrs.fr\\
30.&Nemenov Leonid (CERN \& Dubna)&leonid.nemenov@cern.ch\\
31.&Nyffeler Andreas (Marseille)&nyffeler@cpt.univ-mrs.fr\\
32.&Pancheri Giulia (Frascati)&pancheri@lnf.infn.it\\
33.&Pentia Mircea (Bucharest)&pentia@ifin.nipne.ro\\
34.&Rasche G\"{u}nther (Z\"{u}rich)&rasche@physik.unizh.ch\\
35.&Rusetsky Akaki (Bern \& Tbilisi)&rusetsky@itp.unibe.ch\\
36.&Sainio Mikko (Helsinki)&mikko.sainio@helsinki.fi\\
37.&Santamarina Cibr\'{a}n (Santiago)&cibran.santamarina.rios@cern.ch\\
38.&Sazdjian Hagop (Orsay)&sazdjian@ipno.in2p3.fr\\
39.&Schacher Juerg (Bern)&schacher@lhep.unibe.ch\\
40.&Schumann Marc (Basel)&marc.schumann@unibas.ch\\
41.&Sch\"{u}tz Christian (Basel)&christian.schuetz@cern.ch\\
42.&Schweizer Julia (Bern)&schweize@itp.unibe.ch\\
43.&Simons Leopold (PSI)&leopold.simons@psi.ch\\
44.&Tarasov Alexander (Dubna)&avt@mpimail.mpi-hd.mpg.de\\
45.&Tauscher Ludwig (Basel)&ludwig.tauscher@cern.ch\\
46.&Trautmann Dirk (Basel)&dirk.trautmann@unibas.ch\\
47.&Voskresenskaya Olga (Dubna)&voskr@mpimail.mpi-hd.mpg.de\\
48.&Zemp Peter (Bern)&zemp@itp.unibe.ch\\
49.&Zmeskal Johann (Vienna)&zmeskal@amuon.imep.univie.ac.at
\end{tabular}

\end{center}

}


\newpage


\section{Scientific program}

\vskip.5cm

\begin{itemize}

\item[]\hfill  {\bf Page}

\vskip.5cm

\item[]
{\bf C. Santamarina}\\
DIRAC (PS-212) experiment at CERN \hfill 6

\item[]
{\bf L. Afanasyev}, A. Tarasov and O.Voskresenskaya\\  
Spectrum of pions from $\pi^+\pi^-$-atom breakup (Born and Glauber
approximations)\hfill 7

\item[]
{\bf L.L. Nemenov}\\              
Possibility to measure the energy splitting between $2S$ and $2P$ states in
$\pi^+\pi^-$ atoms\hfill 8

\item[]
J. Gasser, I. Mgeladze, {\bf A. Rusetsky} and I. Scimemi\\
Hadronic atoms: review of theoretical aspects\hfill 9

\item[]
{\bf V.E. Lyubovitskij}\\   
Spectrum and decays of $\pi^+\pi^-$ and $\pi^- p$ atoms\hfill 10

\item[]
{\bf G. Rasche}\\      
Electromagnetic corrections to the lifetime of pionium\hfill 11

\item[]
{\bf E. Lipartia}, V. Lyubovitskij and A. Rusetsky\\      
Hadronic potentials from effective field theory\hfill 12

\item[]
{\bf T.A. Heim}, K. Hencken, M. Schumann, D. Trautmann and 
G. Baur\\         
Distribution of pions from breakup of pionium\hfill 13

\item[]
{\bf M. Schumann}, T. Heim, K. Hencken, D. Trautmann and 
G. Baur\\     
Higher order corrections to bound-bound excitation cross sections of
pionium\hfill 14

\item[]
T. Heim, {\bf K. Hencken}, M. Schumann, D. Trautmann and 
G. Baur\\      
Calculation of electromagnetic breakup of pionium\hfill 15

\item[]
{\bf A. Lanaro}\\
$\pi K$ atom: Observation and lifetime measurement with DIRAC\hfill 16

\item[]
{\bf H. Sazdjian}\\     
Can one extract $m_ s/\hat m$ from $\pi K$ atom properties?\hfill 17

\item[]
B.~Ananthanarayan, {\bf P. B\"{u}ttiker} and B. Moussallam\\
$\pi K$ scattering and dispersion relations\hfill 18

\item[]
{\bf B. Kubis} and U.-G. Mei\ss ner\\
Isospin violation in pion-kaon scattering\hfill 19

\item[]
{\bf C. Curceanu (Petrascu)}\\          
The first measurement of kaonic nitrogen with DEAR at DAFNE\hfill 20

\item[]
{\bf P. Hasenfratz}\\
The quark condensate on the lattice\hfill 21

\item[]
{\bf G. Colangelo}\\
Chiral and lattice calculations of the $\pi\pi$ scattering lengths\hfill 22

\item[]
G. Colangelo, {\bf J. Gasser} and H. Leutwyler\\
The quark condensate from $K_{e4}$ decays\hfill 23

\item[]
{\bf M. Knecht}\\
Radiative corrections to semileptonic decays\hfill 24

\item[]
{\bf V. Cuplov}\\
Isospin violation in $K_{\ell 4}$ decays\hfill 25

\item[]
M. Knecht and {\bf A. Nyffeler}\\
Resonance estimates of low-energy constants in chiral Lagrangians 
and\\ QCD short-dis\-tan\-ce constraints\hfill 26

\item[]
{\bf M. Hennebach}\\
The new pionic hydrogen experiment at PSI\hfill 27

\item[]
{\bf T. Jensen} and V.E. Markushin\\
Atomic cascade in hadronic atoms\hfill 28

\item[]
{\bf M.E. Sainio}\\
Pion-nucleon analysis\hfill 29

\item[]
N. Fettes and {\bf U.-G. Mei\ss ner}\\
Theory of low-energy pion-nucleon scattering\hfill 30

\end{itemize}


\newpage

\setcounter{equation}{0}
\setcounter{figure}{0}

\begin{center}
{\Large{\bf DIRAC (PS-212) experiment at CERN}}

\bigskip

{\bf C. Santamarina$^\dagger$~\footnote{Now at Basel University.} (on
behalf of DIRAC collaboration)}\\[2mm]

$^\dagger$ {\em Departamento de F\'{\i}sica de Part\'{\i}culas,
Universidade de Santiago de Compostela, Campus Universitario Sur, 15782-
Santiago de Compostela, Spain.}

\end{center}

The theoretical results in the low energy regime of hadronic physics have
reached an accuracy level which should be tested with new and precise
experimental measurements. In particular, a prediction of $2.9\pm 0.1
\times 10^{-15}$ seconds [1] has been made for the lifetime of the ground
state of pionium,
the $\pi^+ \pi^-$ electromagnetic atom, within the framework of 
Chiral Perturbation Theory.

DIRAC-PS 212 [2] is a fixed target experiment, currently running at CERN
PS, which aims to measure the pionium lifetime with a $10\%$ accuracy
analyzing the low relative momentum region of $\pi^+ \pi^-$ pairs spectrum
($Q<25 MeV/c$) resulting from proton target collisions [3].

The DIRAC spectrometer is a double arm telescope with a central magnet. It
is optimized to detect events with two oppositely charged particles with low
relative
momentum. The detectors provide an efficient particle identification to
exclude events with proton, electron or muon contamination.

The evidence of pairs from atomic break-up was first found in Summer
2000. The analysis of 2000 and 2001 data has given us around 5600 atoms
mainly with a Titanium (1800 atoms) 175 $\mu m$ and a Nickel (3600 atoms)
$94$ $\mu m$ targets.

With the analyzed data a preliminary result, considering statistical
errors only, has been obtained both with Nickel and Titanium data. The
combined analysis of both samples gives us a lifetime of
$3.6^{+0.9}_{-0.7}\times 10^{-15} s$, however a detailed analysis of
systematic errors is yet to be performed.

\begin{table}[h]
\begin{center}
\begin{tabular}{|c|c|c|}\hline
  Sample    &        Lifetime       & Statistical \\
Target/Year & ($\times 10^{-15} s$) &    error    \\ \hline
  Ni 2000   &  $2.8^{+1.1}_{-0.8}$  & $\pm 34\%$  \\ \hline
  Ti 2000/1 &  $5.4^{+1.5}_{-1.3}$  & $\pm 26\%$  \\ \hline
   Ti+Ni    &  $3.6^{+0.9}_{-0.7}$  & $\pm 22\%$  \\ \hline
\end{tabular}
\caption{Preliminary results with 2000 and 2001 data. Only
statistical errors have been analyzed.}
\end{center}
\end{table}

\bigskip

{\small

\begin{itemize}

\item[{[1]}]
J.~Gasser, V.E.~Lyubovitskij, A.~Rusetsky and A.~Gall, Phys.
Rev. D {\bf 64} (2001) 016008.

\vsp

\item[{[2]}]
B.~Adeva {\em et al.}, CERN-SPSLC-95-1 ; SPSLC-P-284 (1994).

\vsp

\item[{[3]}]
L.L.~Nemenov, Yad. Fiz. {\bf 41} (1985) 980; Sov. J. Nucl.
Phys. {\bf 41} (1985) 629.

\end{itemize}

}

\newpage

\setcounter{equation}{0}
\setcounter{figure}{0}\setcounter{table}{0}

\begin{center}
  {\Large{\bf Spectrum of pions from $\pi^+\pi^-$-atom breakup \\
      (Born and Glauber approximations)}}

\bigskip

{\bf \underline{L. Afanasyev$^1$}, 
A. Tarasov$^{1,2}$ and O.Voskresenskaya$^1$ }\\[2mm]  

$^1$ {\em Joint Institute for Nuclear Research, 
141980 Dubna, Moscow Region, Russia}

$^2$ {\em Institute for Theoretical Phisics, University of
  Heidelberg, Philosophenweg 19, D-69120, Heidelberg, Germany}
\end{center}

Momentum and angular distributions in $\pi^+\pi^-$-pairs from the
$\pi^+\pi^-$-atom ($A_{2\pi}$) breakup at the electromagnetic
interactions with target atoms have been considered in the Born and
Glauber approximations. Exact analytical expressions
for the spectra for the initial states with orbital quantum number
$l=0$ ($n$S states) have obtained in the first Born approximation.
\begin{center}
\includegraphics[width=\textwidth]{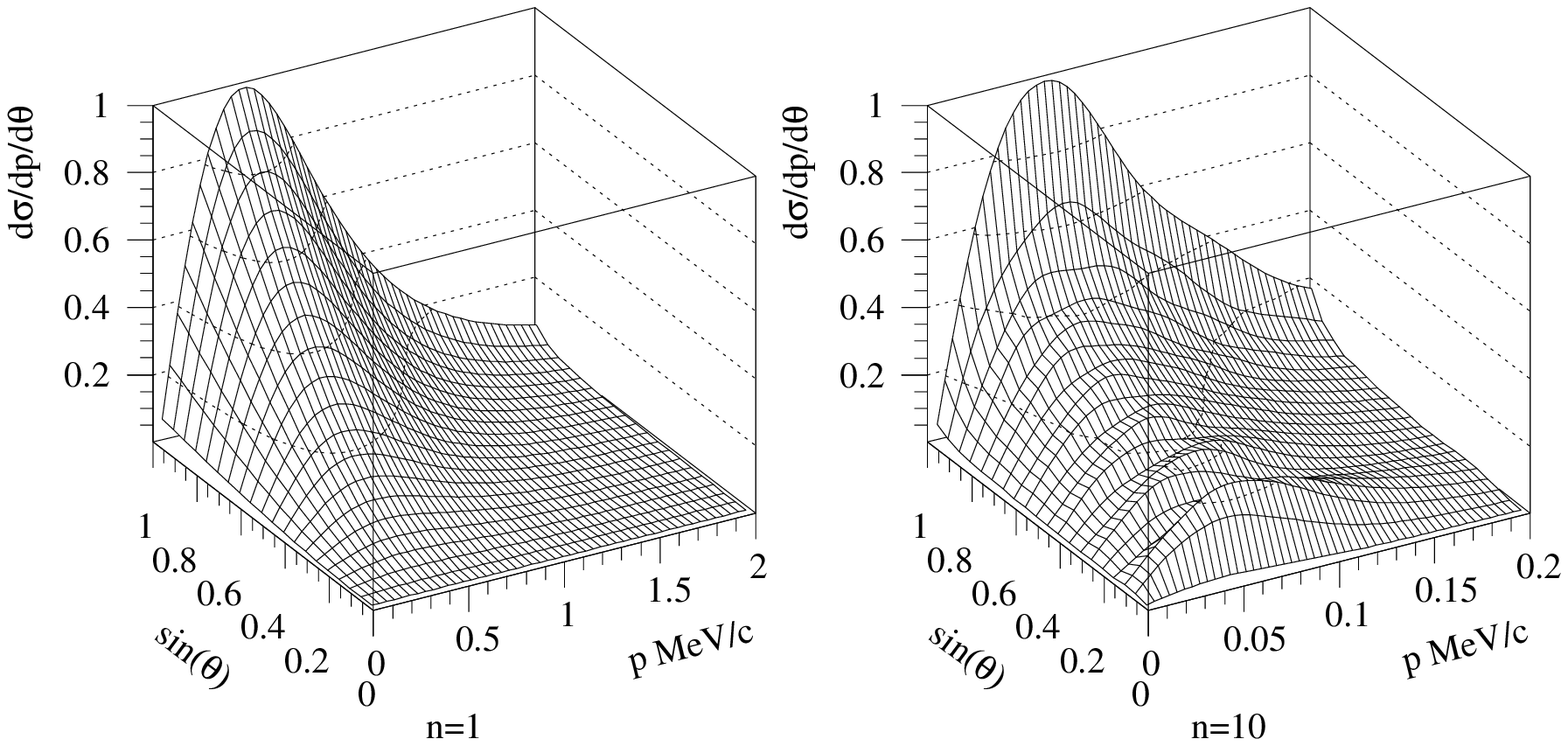}
\end{center}
In the figure the distributions over the relative momentum of pions $p$
and angle $\theta$ with respect to the total momentum of atom are
shown for the initial states with $n=1$ and $n=10$ for the nickel target.

The formulas for the discussed spectra for the initial ground states
of $A_{2\pi}$ which takes into account muli-photon exchange processes
have been obtained in the following approach.  It is shown that due to
the small size of $A_{2\pi}$ and its big mass the interaction with an
unscreened target nuclear dominantly contributes to the multi-photon
exchange processes. The amplitude of the atom interaction with
unscreened Coulomb potential can be expressed in a closed analytical
form in the Glauber approximation. Thus one can consider the target
atom as the screened Coulomb potential for accounting the single-photon
exchange processes in the Born approximation and as unscreened one for
multi-photon processes.

\newpage

\setcounter{equation}{0}
\setcounter{figure}{0}\setcounter{table}{0}

\begin{center}
{\Large{\bf Possibility to measure the energy splitting between 2S \\
\vspace*{0.1cm} 
and 2P states in $\pi^+ \pi^-$ atoms}}

\bigskip

{\bf L.L. Nemenov}\\[2mm]  

{\em CERN, CH-1211, Geneva 23, Switzerland and Dubna (JINR), Russia} 

\end{center}

The difference $\Delta E_n$ between the energy of atomic $nS$ and $nP$ states
of $A_{2\pi}$ includes the contributions of the vacuum polarisation 
$\Delta E_n^{vac}$, and of the strong interaction $\Delta E_n^{s}$ effects,
where $\Delta E_n^{s} \sim 2a_0+a_2$, and $a_0$ and $a_2$ are the S-wave
$\pi \pi$ scattering lengths with isotopic-spin quantum numbers 0,2 [1,2,3].

The value of $\Delta E_n^{vac}$ is well known from QED calculations [1,2,4]. 
For this reason the measurement of $\Delta E_n$ will give, in a model 
independent way, the value of $2a_0+a_2$ [5,6].

The method for measuring $\Delta E_n$ is presented [7,8]. The same method 
allows
also to measure the energy splitting ($\Delta E_n$) for atoms consisting of 
$\pi$
and $K$ mesons, and to obtain the value of $2a_{1/2}+a_{3/2}$, where 
$a_{1/2}$
and $a_{3/2}$ are the S-wave $\pi K$ scattering lengths with isotopic-spin
1/2 and 3/2, respectively.

\bigskip

{\small

\begin{itemize}

\item[{[1]}]
A.~Gashi et al., Nucl. Phys. A {\bf 628} (1999) 101.

\vsp

\item[{[2]}] 
A.~Rusetsky, private communication.

\vsp

\item[{[3]}]
G.~Efimov et al., Sov. J. Nucl. Phys. {\bf 44} (1986) 296.

\vsp

\item[{[4]}]
A.~Karimkhodzhaev and R.N.~Faustov, Sov. J. Nucl. Phys. {\bf 29} (1979) 232.

\vsp

\item[{[5]}]
G.~Colangelo, J.~Gasser and H.~Leutwyler, Phys. Lett. B {\bf 488} (2000) 261.

\vsp

\item[{[6]}]
M.~Knecht et al., Nucl. Phys. B {\bf 457} (1995) 513.

\vsp

\item[{[7]}]
L.L.~Nemenov, Sov. J. Nucl. Phys.{\bf 41} (1985) 629.

\vsp

\item[{[8]}]
L.L.~Nemenov and V.D.~Ovsiannikov, Phys. Lett. B {\bf 514} (2001) 247. 
 
\end{itemize}
}

\newpage

\setcounter{equation}{0}
\setcounter{figure}{0}\setcounter{table}{0}

\begin{center}
{\Large{\bf Hadronic atoms: review of theoretical aspects}}

\bigskip

{\bf J. Gasser, I. Mgeladze, \underline{A. Rusetsky} and I. Scimemi}\\[2mm]  

{\em Institute for Theoretical Physics, University of Bern,
Sidlerstrasse 5, CH-3012, Bern, Switzerland}

\end{center}

In order to fully exploit the high-precision experimental data on
hadronic atoms which is available at present and which will be provided in
the future, it is
imperative to design a theoretical framework that describes this  type
of bound systems to the accuracy that
matches the experimental precision. 
The Deser-type relations which are then used  to extract the ``purely 
strong''
scattering lengths from the measured values of the strong energy shift 
in the ground state, and the partial decay width into 
hadronic channels, contain isospin-breaking corrections which depend on  
the fine structure constant $\alpha$ and on
the  quark mass difference $m_d-m_u$. The following questions
 arise.
\begin{itemize}
\item[i)]
The scattering lengths are calculated in  QCD at $m_u=m_d$.
 How is this theory related to a framework that includes
isospin violating effects,  $SU(3)_c \times U(1)_{em}$?
[Note that, in view of the announced accuracy $0.3\,\%$ 
in the future measurement of the energy shift of the $\pi^-p$ atom by  
Pionic Hydrogen collaboration at PSI, the question of the precise
definition of the pure QCD limit is not an academic one.]
\item[ii)]
With a given definition of the pure QCD limit, how does one calculate
the isospin-breaking corrections to the bound-state observables?
\end{itemize}

The answers are as follows.

\begin{itemize}
\item[{ad i)}]
In two-flavour QCD, the quark mass $m_u=m_d$ and the strong coupling
constant may be  fixed from the requirement that 
the pion and nucleon masses are
equal - by convention - to the experimental values of the charged pion 
and the proton masses, respectively. In the effective theory 
of $SU(3)_c\times U(1)_{em}$, the values of the strong LECs stay put [1].
\item[{ad ii)}]
We have developed a general theory of hadronic atoms,
based on a merger of non-relativistic effective Lagrangian
techniques and ChPT~[3-6]. This theory has been successfully applied
to the calculation of the isospin-breaking corrections in the
$\pi^+\pi^-$ atom decay width~[3,4,6], and the energy shift of the
$\pi^-p$ atom~[5]. For related work, see [7]. Applications to other 
systems are in preparation [8].
\end{itemize}

\bigskip

{\small

\begin{itemize}

\item[{[1]}]
J. Gasser, I. Mgeladze, A. Rusetsky and I. Scimemi, in preparation.
This reference reviews earlier work on the subject. We have profited
from unpublished notes by H.~Leutwyler, whom we thank in addition for
informative discussions on the subject.

\vsp

\item[{[3]}]
A.~Gall, J.~Gasser, V.E.~Lyubovitskij and A.~Rusetsky, 
Phys.\ Lett.\ B {\bf 462} (1999) 335.

\vsp

\item[{[4]}]
J.~Gasser, V.E.~Lyubovitskij and A.~Rusetsky,
Phys.\ Lett.\ B {\bf 471} (1999) 244 (1999).

\vsp

\item[{[5]}]
V.E. Lyubovitskij and A. Rusetsky, 
Phys.\ Lett.\ B {\bf 494} (2000) 9.

\vsp

\item[{[6]}]
J.~Gasser, V.E.~Lyubovitskij,  A.~Rusetsky and A.~Gall, 
Phys.\ Rev.\ D {\bf 64} (2001) 016008.

\vsp

\item[{[7]}]
X. Kong and F. Ravndal, Phys.\ Rev.\ D 61 (2000) 077506 ; 
D. Eiras and J. Soto, Phys.\ Lett.\ B 491 (2000) 101, and references
cited therein.

\vsp

\item[{[8]}]
M.~Schmid, J.~Schweizer and P.~Zemp, work in progress.

\end{itemize}
}

\newpage

\setcounter{equation}{0}
\setcounter{figure}{0}\setcounter{table}{0}

\begin{center}
{\Large{\bf Spectrum and decays of $\pi^+\pi^-$ and $\pi^- p$ atoms [1]}}

\bigskip

{\bf V. E. Lyubovitskij} 

{\em Institut f\"ur Theoretische Physik, Universit\"at T\"ubingen, \\ 
Auf der Morgenstelle 14, D-72076 T\"ubingen, Germany} 

\end{center}

In order to carry out the precision experimental tests of QCD 
we proposed and developed {\it effective theory} [2,3] for 
the description of measured characteristics of hadronic atoms. 
We have derived the expressions for the decay width $\Gamma_{2\pi^0}$ 
of the $\pi^+\pi^-$ atom [2] and strong energy shift of the 
$\pi^- p$ atom [3] in the framework of QCD (including photons) by use 
of effective field theory techniques. The results obtained contains all 
terms at leading and next-to-leading order in the isospin breaking 
parameters $\alpha\simeq 1/137$ and $m_u - m_d$. On the basis 
of these formulae, a numerical analysis was carried out in 
Refs. [2,3] at order $e^2p^2$ in chiral perturbation theory (ChPT). 

The $\pi^+\pi^-$ atom decay problem is completely solved [2]. 
Numerical analysis in ChPT based on recent update results 
for the $S$-wave $\pi\pi$ scattering lengths $a_0$ and $a_2$ gives 
\begin{eqnarray}
\tau\doteq \Gamma_{2\pi^0}^{-1}=(2.9\pm 0.1)\times10^{-15}~{\mbox s} \, .
\end{eqnarray}
Numerical analysis of the $\pi^- p$ atom strong energy shift is in 
progress because the effective electromagnetic coupling constant 
$f_1$ is still unknown. 

The perturbative chiral quark model [4] was applied 
to estimate the electromagnetic $O(p^2)$ low-energy constants of the 
ChPT effective Lagrangian that define the electromagnetic mass shifts of 
nucleons and first-order $(e^2)$ radiative corrections to the $\pi N$ 
scattering amplitude [5]. It was found that the value of $f_2$ constant 
is consistent with model-independent result obtained by Gasser and 
Leutwyler [6]. Using obtained value for constant $f_1$ the leading 
isospin-breaking correction to the strong energy shift of the $\pi^- p$ 
atom in the $1s$ state was calculated. The result 
$\delta_{\epsilon} = - 2.8 \times 10^{-2}$ is comparable to a prediction 
based on a potential model for the $\pi N$ scattering~[7]: 
$\delta_{\epsilon} = - 2.1 \times 10^{-2}$.

{\it Acknowledgements}. This work was supported by the Deutsche
Forschungsgemeinschaft (DFG, grant FA67/25-1).

\bigskip

{\small

\begin{itemize}

\item[{[1]}]
Work done in collaboration with J.~Gasser, A.~Rusetsky, 
A.~Gall [2,3] and Th.~Gutsche, A.~Fa\-ess\-ler, R.~Vinh Mau [4,5].  

\vsp

\item[{[2]}]
A.~Gall, J.~Gasser, V.E.~Lyubovitskij and A.~Rusetsky, 
Phys. \ Lett. \ B {\bf 462} (1999) 335;
J.~Gasser, V.E.~Lyubovitskij and A.~Rusetsky, 
Phys. \ Lett. B {\bf 471} (1999) 244;
J.~Gasser, V.E.~Lyubovitskij, A.~Rusetsky and A.~Gall, 
Phys. \ Rev. D {\bf 64} (2001) 016008.  

\vsp

\item[{[3]}]
V.E.~Lyubovitskij and A.~Rusetsky, 
Phys. \ Lett. B {\bf 494} (2000) 9. 

\vsp

\item[{[4]}]
V.E.~Lyubovitskij, T.~Gutsche, A.~Faessler and E.G.~Drukarev, 
Phys. Rev. D {\bf 63} (2001) 054026. 

\vsp

\item[{[5]}]
V.E.~Lyubovitskij, T.~Gutsche, A.~Faessler and R.~Vinh Mau, 
Phys. Lett. B {\bf 520} (2001) 204. 

\vsp

\item[{[6]}]
J.~Gasser and H.~Leutwyler,
Phys. \ Rep. {\bf 87} (1982) 77. 

\vsp

\item[{[7]}]
D.~Sigg, et al, Nucl. Phys. A {\bf 609} (1996) 310. 
\end{itemize}

}

\newpage

\setcounter{equation}{0}
\setcounter{figure}{0}\setcounter{table}{0}

\begin{center}
{\Large{\bf Electromagnetic corrections to the lifetime of pionium}}

\bigskip

{\bf G. Rasche}\\[2mm]  

 {\em Institute for Theoretical Physics, University of Z\"{u}rich,
Winterthurerstr. 190, CH-8057, Z\"{u}rich, Switzerland}

\end{center}

We have made use of the Deser formula for the two-channel system as derived 
in [1].  Including a tiny correction for vacuum polarisation etc. in the
numerical constant, we get from [2]: 
\begin {equation}
\tau\, (\rm fs) = \frac {0.09474} {({\it a}_{0c}\, (\rm fm))^2}\,\,.
\end {equation}
Here $a_{0c}$ is the value of the $\pi^+\pi^-\rightarrow\pi^0\pi^0$ 
transition element of the  $\bf{K}$-matrix of scattering theory at the  
$\pi^+\pi^-$ threshold. It consists of a hadronic part $a^h_{0c}(\mu_0)$ and
an electromagnetic correction $\Delta a_{0c}(\mu_0)$: 
\begin {equation}
a_{0c}=a^h_{0c}(\mu_0)+\Delta a_{0c}(\mu_0)\,\,.
\end {equation}

Since the pion mass for the hadronic limit is very close to the neutral pion
mass $\mu_0$ it follows that the hadronic starting point for the true
electromagnetic corrections is  $\mu_0$. We have indicated this explicitly on
the lhs of (2).

The value for  $a^h_{0c}(\mu_0)$ is taken from [3]; they use $a^0-a^2 =
0.265(4)$.  Taking into account that this value corresponds to a hadronic
starting point with pion mass $\mu_c$ and is given in units of $\mu_c^{-1}$
one gets (see [2]) 
\begin {equation}
 a^h_{0c}(\mu_0)=\frac{\sqrt{2}}{3}(a^2(\mu_0)-a^0(\mu_0))=-0.170(3)\, \rm fm
 \, 
\,.
\end {equation}
The electromagnetic correction  $\Delta a_{0c}(\mu_0)$ is calculated using a
relativised two-channel Schr\"{o}\-din\-ger equation with two isospin 
invariant
energy independent potentials $V^I(r), I=0,2$. They are constructed by
fitting an analytical  ansatz to the two-loop hadronic ChPT scattering phase
$\delta^{I}(q), I=0,2$ of [4], for both values $\mu_0$ and $\mu_c$ of the 
pion mass. Solving the  coupled Schr\"{o}dinger equation including the 
Coulomb interaction and the mass difference $\mu_c-\mu_0$, as well as taking
into account the results of [5], we arrive at 
\begin {equation}
\Delta a_{0c}(\mu_0)=0.010(1)\, \rm fm\,\,.
\end {equation}
Inserting these results into (1) we get
\begin {equation}
\tau=2.92(15)\, \rm fs\,\,.
\end {equation}
This agrees with [3].

\bigskip

{\small

\begin{itemize}

\item[{[1]}]
G. Rasche and W.S. Woolcock, Nucl.Phys. A {\bf 381} (1982) 405.

\vsp
 
\item[{[2]}]
A. Gashi, G.C. Oades, G. Rasche and W.S.Woolcock, arXiv:hep-ph/0108116, to 
appear in Nucl. Phys. A. 

\vsp

\item[{[3]}]
J. Gasser, V.E. Lyubovitskij, A. Rusetsky and A. Gall, Phys. Rev. D {\bf 64}
(2001) 016008. 

\vsp

\item[{[4]}]
J. Bijnens, G. Colangelo, G. Ecker, J. Gasser and M.E. Saino, Phys.Lett. B
{\bf 374}
(1996) 210; Nucl. Phys. B {\bf 508} (1997) 263 and erratum ibid. B
{\bf 517}  (1998) 693; 
G. Colangelo, private communication.

\vsp

\item[{[5]}]
M. Knecht and R.Urech, Nucl. Phys. B {\bf 519} (1998) 329.

\end{itemize}

}

\newpage

\setcounter{equation}{0}
\setcounter{figure}{0}\setcounter{table}{0}

\begin{center}
{\Large{\bf Hadronic potentials from effective field theory}}

\bigskip

{\bf \underline{E. Lipartia}$^{1,2,3}$, V. Lyubovitskij$^{4,5}$ 
and A. Rusetsky$^{3,6}$}\\[2mm]  

$^1$ {\em \small{Department of Theoretical Physics 2, Lund University, 
Solvegatan 14A, S22362 Lund, Sweden}}

$^2$ {\em \small{ LIT, Joint Institute for Nuclear Research, 141980, Dubna, 
Russia}}

$^3$ {\em \small {HEPI, Tbilisi State University, University st. 9, 
380086 Tbilisi, Georgia}}

$^4$ {\em \small{Institute of Theoretical Physics, University of T\"ubingen,
Auf der Morgenstelle 14, D-72076, T\"ubingen, Germany}}  

$^5$ {\em \small{Department of Physics, Tomsk State University, 634050 
Tomsk, Russia}}

$^6$ {\em\small{ Institute for Theoretical Physics, University of Bern,
Sidlerstrasse 5, CH-3012, Bern, Switzerland}}

\end{center}

The problem of the evaluation of isospin symmetry breaking effects to the
hadronic 
atom observables has been addressed in the framework of field-theoretical 
approaches, as well as in the potential models. At the same time, 
the results obtained within 
the potential model do not always agree with the results of 
field-theoretical 
approaches. The main reason of discrepancy between the predictions of both 
approaches is that, in general, the potential model does not take into 
account the full 
content of  isospin-breaking effects in ChPT. 
Even if these predictions agree in a particular case 
(e.g., like the latest prediction of the potential 
model for the lifetime of $\pi^+\pi^-$ atom~[1]), without an explicit
matching to ChPT in the isospin-breaking phase the treatment of the problem 
in the potential model can not be regarded as systematic.
In the view of the 
fact that the potential approach is widely used to evaluate the
isospin-breaking effects in scattering process - to analyze the results of 
experimental measurements - it is very important to construct the potential
which includes the full content of isospin-breaking effects.

The constructive algorithm for the derivation of the hadronic potential from
ChPT is considered in detail in our paper~[2]. Here we provide only a brief
outline of the procedure.
The derivation of the isospin-breaking part of the short-range strong
potential is based on the universality conjecture, which states that the 
relation between the bound-state characteristics and scattering amplitudes is
the same in the field theory and in the potential model. The universality 
conjecture provides us with the matching condition for the isospin-breaking 
part
of hadronic potential, assuming the isospin-symmetric part of the potential 
is given. Further, the matching condition imposes rather loose constraints on
the potential. This looseness reflects the freedom of the choice 
of regularization of pointlike singularities which emerges if the potential
is derived from a local field theory. For
this reason, the shape of the potential does not bear a physical information,
and using this freedom, one may absorb all information about the 
isospin-breaking effects in the couplings of the short-range potential which,
in its turn, are determined from the matching condition~[2].

Despite the
unrestricted freedom in the choice of the shape of 
the short-range part of the potential, the full potential that we 
construct,
reproduces, by construction, the threshold 
amplitude and the bound-state characteristics of hadronic atoms at the first
non-leading order in isospin-breaking.

\bigskip

{\small

\begin{itemize}

\item[{[1]}]
G. Rasche, these proceedings.

\vsp

\item[{[2]}]
E.~Lipartia, V.E.~Lyubovitskij and A.~Rusetsky, arXiv:hep-ph/0110186.

\end{itemize}

}

\newpage

\setcounter{equation}{0}
\setcounter{figure}{0}\setcounter{table}{0}

\begin{center}
{\Large{\bf Distribution of pions from breakup of pionium}}

\bigskip

{\bf \underline{T.A. Heim}$^1$, K. Hencken$^1$, 
M. Schumann$^1$, D. Trautmann$^1$ 
and G. Baur$^2$}\\[2mm]  

$^1$ {\em Institut f\"ur Physik, Universit\"at Basel, 
Klingelbergstrasse 82, CH--4056, Basel, Switzerland}

$^2$ {\em Institut f\"ur Kernphysik, Forschungszentrum J\"ulich, 
52425 J\"ulich, Germany}
\end{center}
The experiment DIRAC requires in its analysis a reliable calculation for
the number of pionium atoms produced in the target.
At small relative momentum $Q<3$MeV/$c$ the observed momentum distribution
shows an excess of $\pi^+\pi^-$ pairs (attributable to `atomic' pairs)
as compared to the momentum 
distribution obtained from a fit at larger relative momentum where only
`free' pairs contribute [1]. However, extracting the number of pionium
atoms in this way requires accurate (momentum) distributions of pions from
the breakup of pionium as an input. 

In the framework of the semiclassical approximation [2], 
the distribution of
the pions from the breakup, resolved with respect to magnitude and  
angular direction of the momentum of the out-coming pions, is given by
\begin{displaymath}
\frac{\mathrm{d}^2\sigma_i}{\mathrm{d}\Omega\,\mathrm{d}p} =
\int\mathrm{d}^2b\,
\left|\sum_{l_f,m_f}a_{fi}(b)
Y_{l_f,m_f}(\Omega)\right|^2,
\end{displaymath}
where $a_{fi}(b)$ denotes the amplitude for the transition from the
bound initial state $i$ to the final continuum state with (internal)
relative momentum $p$ of the pionium. Here,  
$p=m_{\mathrm{red}}\dot{r}=\frac12(p_+-p_-)$, with $p_{\pm}$ 
denoting the momentum of the two pions 
in their center-of-mass frame. With the method outlined in [2] we can
calculate these transitions into the continuum directly for arbitrary
initial states.  
The azimuthal symmetry with respect to
the beam axis reduces the angular distribution to a mere 
$\theta$-dependence where $\theta$ denotes the angle between the beam and
the out-coming $\pi^+$ (or $\pi^-$, since the distribution is 
symmetric under the reflection $\theta\leftrightarrow\pi-\theta$ in 
first order of the interaction, i.e., considering only breakup of the 
pionium due to one-photon exchange and taking into account the nearly
perfect $z$-parity conservation [3,4], which becomes exact in the sudden 
limit).

We have calculated and analyzed the differential cross sections for breakup
of pionium on a Nickel target at projectile energy 5~GeV, for 
initial states $n_i$s and $n_i$p with $n_i\le4$. 
We found that when integrated over the
momentum, the angular distribution shows the two pions to be 
preferentially emitted perpendicular to the beam, as expected. This strongly
non-isotropic distribution
is to be contrasted with the isotropic distribution of the `free' pairs
not stemming from the breakup of pionium. 

When integrated over the angles, the momentum distribution for breakup from
an initial state with principal quantum number $n_i$ has its peak at
$p\approx (1/2n_i)$~MeV/$c$, with a width of approximately
$\Delta p\approx (1/2n_i)$~MeV/$c$. The peak amplitude scales as $n_i^2$, as 
expected. Similar calculations for other initial states (d- and f-states
reached by multiple scattering) and other target materials will be 
carried out. 
These momentum distributions will then be used in a more accurate extraction
of the number of pionium atoms produced in the experiment. 

\bigskip

{\small

\begin{itemize}

\item[{[1]}]
L.G. Afanasyev \emph{et al.}, Phys. Lett. B \textbf{338} (1994) 478.

\vsp

\item[{[2]}]
Z. Halabuka \emph{et al.}, Nucl. Phys. B \textbf{554} (1999) 86.

\vsp

\item[{[3]}]
T.A. Heim, K. Hencken, D. Trautmann and G. Baur, J. Phys. B \textbf{33}
(2000) 3583.

\vsp

\item[{[4]}]
T.A. Heim, K. Hencken, D. Trautmann and G. Baur, J. Phys. B \textbf{34}
(2001) 3763.

\end{itemize}

}

\newpage

\setcounter{equation}{0}
\setcounter{figure}{0}\setcounter{table}{0}

\begin{center}
{\Large{\bf Higher order corrections to the bound-bound excitation cross
sections of pionium}}\\
\bigskip
{\bf \underline{M.\ Schumann}$^1$, T.\ Heim$^{1}$, K.\ Hencken$^{1}$, 
D.\ Trautmann$^{1}$ and G.\ Baur$^2$}\\[2mm]  
$^1$ {\em Institut f\"ur Physik, Universit\"at Basel,
Klingelbergstr.\ 82, CH-4056, Basel, Switzerland}\\
$^2$ {\em Institut f\"ur Kernphysik, Forschungszentrum J\"ulich, 
52425 J\"ulich, Germany}
\end{center}
The total and excitation cross sections for the interaction of pionium with 
the
target atoms are an essential theoretical input for the experiment DIRAC [1].
They are well known in the Born approximation and can be calculated to a very
high accuracy including various corrections [2--4]. However, Afanasyev et 
al.~[5] showed that the total cross section in Born approximation 
overestimates
the value in Glauber approximation by up to 14\% for $Z=73$ (Ta). This is a
major setback for the prospect of calculating the cross sections to 1\% as is
required for the success of the experiment. To calculate the excitation cross
sections we start with the transition amplitude between an initial state $i$
and a final state $f$ given by \vspace{-0.2cm}
$$
 a_{fi}^G(b)=\int d^3\!r\, \psi_f^\ast(\vec r)
\left[ 1-\exp\mbox{\boldmath $($}i\chi(b,\vec r)\mbox{\boldmath $)$}\right]
                \psi_i(\vec r)\ ,\vspace{-0.2cm}
$$
where $\psi_{i,f}(\vec r)$ are non-relativistic hydrogen-like wave 
functions and 
$\chi(b,\vec r)=-(e/\hbar v) \int [\Phi(\vec r\,'/2)\\-\Phi(-\vec r\,'/2)] 
dz$ with
the screened potential $\Phi(r')=(Ze/r')\sum A_k \exp(-\alpha_k r')$. 
Unlike in
the case of the total cross section, where the closure approximation leads to
significant simplifications, the excitation cross sections cannot
be reduced to a one-dimensional integral in general. Instead a full numerical
integration is required. Since the cross sections in Born approximation are 
well
known, we only need to calculate the higher order corrections.
Therefore we evaluate the integral \vspace{-0.35cm}
$$\Delta_{fi}=2\pi\int b\,db\,[|a_{fi}^B(b)|^2-|a_{fi}^G(b)|^2]\ ,
\vspace{-0.05cm}$$
which converges much faster than the full cross section 
since the higher order corrections are most important for small impact 
parameters $b$.
Figure~1 shows the results for a few selected transitions. One can see that 
even
for a Nickel target ($Z=28$) the higher order corrections are between 3\% and
5\%, while for large $Z$ they reach up to 30\%. Clearly, we need to take 
these
results into consideration to reach the desired accuracy of 1\% for the 
cross sections. \\
\begin{figure}[thb]
\vspace{-0.5cm}
\hspace{\fill}
\parbox{6cm}{
\includegraphics[width=5cm]{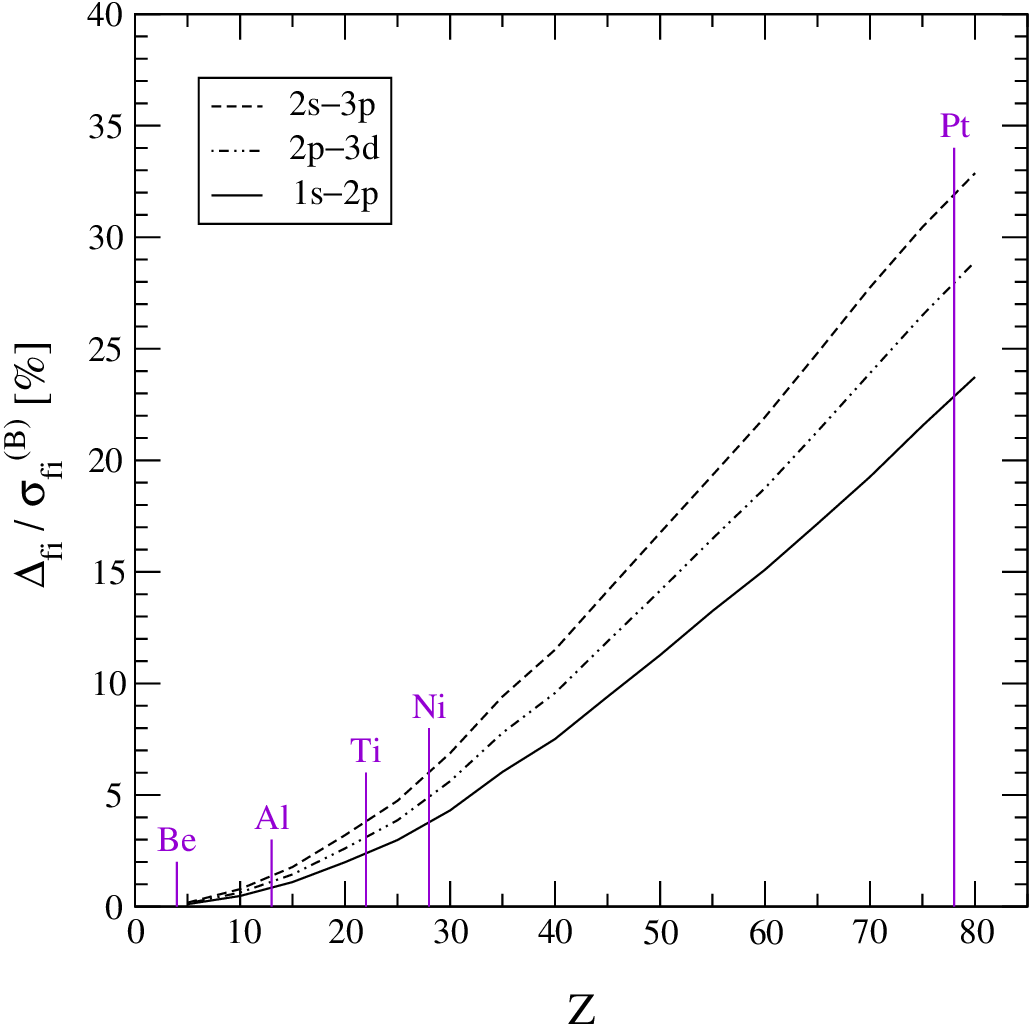}}
\parbox{7cm}{Figure 1: Relative difference between Glauber and Born
approximation for the excitation cross sections of 1s--2p, 2p--3d, and 2s--3p
transitions as a function of the target charge number $Z$.}
\hspace{\fill}
\end{figure}
{\small
\begin{itemize}

\item[{[1]}]
B.~Adeva et al., CERN/SPSLC 95--1, SPSLC/P 284, Geneva (1995).

\vsp

\item[{[2]}]
Z.~Halabuka, et al.,  
Nucl.~Phys.~B {\bf554} (1999) 86.

\vsp

\item[{[3]}]
T.A.~Heim, K.~Hencken, D.~Trautmann, and G.~Baur, 
  J.~Phys.~B {\bf 33} (2000) 3583.

\vsp

\item[{[4]}]
T.A.~Heim, K.~Hencken, D.~Trautmann, and G.~Baur, 
  J.~Phys.~B {\bf 34} (2001) 3763.

\vsp

\item[{[5]}]
L.~Afanasyev, A.~Tarasov, and O.~Voskresenskaya, 
  J.~Phys.~G {\bf 25} (1999) B7.

\end{itemize}
}

\newpage

\setcounter{equation}{0}
\setcounter{figure}{0}\setcounter{table}{0}

\begin{center}
{\Large{\bf Calculation of electromagnetic breakup of pionium}}

\bigskip

{\bf T. Heim$^1$, \underline{K. Hencken}$^1$, M. Schumann$^1$, D. 
Trautmann$^1$ and G. Baur$^2$}\\[2mm]  

$^1$ {\em Institut f\"ur Physik, Universit\"at Basel,
Klingelbergstr. 82, CH-4056, Basel, Switzerland}

$^2$ {\em Institut f\"ur Kernphysik, Forschungszentrum J\"ulich, 
D-52425 J\"ulich, Germany}
\end{center}

The calculation of the electromagnetic excitation and breakup of Pionium in 
the target is an important input for the analysis of the experiment DIRAC at
CERN. The cross section for excitation, as well as, breakup was studied 
within the semiclassical approximation for the dominant Coulomb part of the 
interaction (``scalar interaction'') in [1]. In addition to coherent 
excitations, where the atom remains in the ground state, incoherent 
excitations
have to be considered also. Summing over all excited states using the closure
approximation one obtains a compact expression in terms of the inelastic 
scattering function $S_{inel}$ [2]. Here only the dominant Coulomb 
interaction
is considered. The $S_{inel}$ were calculated within a 
Dirac-Hartree-Fock-Slater calculation. Simpler models like ``antiscreening''
and the ``no-correlation limit'' were found to be not adequate for the 
accuracy
needed. The contribution of transverse photons was estimated to be at most 
0.4\% by making use of the long-wavelength approximation.

Calculation of higher order effects are studied within the Glauber 
approximation and are discussed in [3]. This leaves us now with the 
calculation of the ``magnetic interaction'', that is, the current 
interaction 
as the estimate show it to contribute almost 1\%.
In addition we want to study possible effects due to the relativistic 
treatment
of the interaction between pionium and target [4]. The nonrelativistic 
Schr\"odinger wave 
function for the Pionium states could easily be replaced by the solution 
of the Klein-Gordon equation, but this gives only corrections of order 
$\alpha^2$. Starting from the Klein-Gordon equation, we separate
terms coming from the external field $A_X$, treating them as a perturbation.
Using the Feshbach-Villars reduction to separate ``large'' and 
``small components'', we find the interaction between pionium and target to 
be
$$
i e A_{X0} + \frac{i e}{2 m} \left[ \partial_0 A_{X0} + 
\vec\nabla \vec A_X + \vec A_X \vec\nabla\right] - \frac{e^2}{2 m} 
\left( A_{X0}^2 - {\vec A_X}^2 \right) - \frac{e^2}{m} \Phi_C A_{X0}.
$$
Besides the usual interaction terms, two new contributions are found:
The interaction quadratically in $A$ is given by
$$
A_X^\mu A_{X\mu}= A_{X0}^2 -{\vec A_X}^2 = \frac{1}{\gamma^2} A_0^2,
$$ 
which is suppressed by $1/\gamma^2$. The nonrelativistic treatment only leads
to $\vec A_X^2$ instead, which is not necessarily small. The second term is
a Seagull diagram of the form $\Phi_C A_{X0} $ with $\Phi_C$ the Coulomb 
potential between $\pi^+ \pi^-$. The same terms are also found starting from
a diagrammatic approach, taking into account that Pionium consists of two
equally heavy constituents.
Explicit calculations of both the magnetic term, as well as, the Seagull 
diagram are performed. The Seagull diagram is found to contribute only
to order $10^{-5}$; the magnetic terms are found to be less than 1\%,
in agreement with the estimates.

\bigskip

{\small

\begin{itemize}

\item[{[1]}]
Z. Halabuka, et al., Nucl. Phys. B {\bf 554} (1999) 86.

\vsp

\item[{[2]}]
T. A. Heim, K. Hencken, D. Trautmann, and G. Baur,
J. Phys. B {\bf 33} (2000) 3583.

\vsp

\item[{[3]}]
M. Schumann et al., contribution to this miniproceedings.

\vsp

\item[{[4]}]
T. A. Heim, K. Hencken, D. Trautmann, and G. Baur, J. Phys. B {\bf 34}, 
(2001) 3763.

\end{itemize}
}

\newpage

\setcounter{equation}{0}
\setcounter{figure}{0}\setcounter{table}{0}

\begin{center}
{\Large{\bf $\pi K$ atom: observation and lifetime measurement with \\
\vspace*{0.3cm}
DIRAC}}

\bigskip

{\bf A. Lanaro}\\[2mm]  
{\em (on behalf of the DIRAC Collaboration)}\\[2mm]

{\em CERN, CH-1211, Geneva 23, Switzerland} 

\end{center}

The detection of $\pi K$ atoms ($A_{\pi K}$) and the measurement of their 
lifetime
provide a model independent way to determine $|a_{1/2}-a_{3/2}|$, the
difference  
between I=1/2,3/2 S-wave $\pi K$ scattering lengths. The $\pi K$ 
scattering probes the $SU(3)_L \times SU(3)_R$ predictions of Chiral
Perturbation Theory (ChPT) with $u, d, s$, and unequal mass kinematics,
whereas 
the investigation of $\pi \pi$ scattering checks the symmetry
breaking part of $SU(2)_L \times SU(2)_R$ with $u$ and $d$ quarks only.

The $\pi K$  scattering amplitude has been evaluated at next-to-leading 
order in standard ChPT [1].
On the experimental side, $S$-wave $\pi K$ phase shifts have been measured
from threshold up to 1.3 GeV, using the scattering of moderately energetic 
charged kaon beams ($2<E_K<13$ GeV). The results for the I=1/2 and I=3/2 
S-wave $\pi K$ scattering lengths
are spread over a wide range: $0.13<a_{1/2}<0.24$,
$-0.13<a_{3/2}<-0.05$, in units of inverse $\pi$ mass. 

To determine the $\pi K$ threshold parameters, the DIRAC Collaboration
proposes the same experimental approach as in the 
$\pi \pi$ case, which consists in the detection of $\pi K$ atoms [2].

$A_{\pi K}$ are detected because they have a sizeable
probability to break-up in traversing the target material where they
are produced. The ratio of experimentally detected $\pi^{\pm} K^{\mp}$ atomic
pairs to the full number of produced atoms, yields a measurement of the
break-up probability in a given target, and, thus, of the $A_{\pi K}$
lifetime. The relation between the ground state lifetime and the difference
$|a_{1/2}-a_{3/2}|$ is given, in first order, by the Deser formula [3]. Using
the ChPT prediction at leading order in Isospin breaking for 
$a_{1/2}$ and $a_{3/2}|$ [1], the lifetime of the $A_{\pi K}$ ground state is
predicted to be 4.8 $\cdot$ 10$^{-15}$ s.   
   
The Collaboration envisages to use the present detection apparatus, as well 
as incoming particle
beam and energy (24 GeV/$c$ protons from the CERN PS). Only some new 
implementations to the particle identification system would have to be 
considered to allow the simultaneous detection of charged pions and kaons.
Additional threshold Cherenkov counters of different type have been proposed
[2].  
Two identical detectors will employ a heavy gas radiator (Freon 114), 
suitable to discriminate $\pi^{\pm}$ from  $K^{\pm}$ in the kinematic range 
of accepted $A_{\pi K}$. 
Another Cherenkov detector will employ Aerogel as radiator to 
discriminate $K^+$ from protons.

In the summer of 2001 a test has been performed using the present
Cherenkov detectors filled with SF$_6$ gas. The discrimination of pions
from kaons was investigated by detecting the $\phi(1020) \rightarrow K^+K^-$ 
resonance decay mode. During 2002, the Collaboration will also perform tests 
of Aerogel samples with very low index of refraction (n=1.008),
manufactured by the Boreskov Institute of Catalysis at Novosibirsk.

In three years of running time, DIRAC expects to
detect a few thousands $\pi K$ atomic pairs, which should lead
to a determination of $|a_{1/2}-a_{3/2}|$ with 10$ \div$ 15$\%$ 
accuracy, and to a substantial improvement on the precision of the
$\pi \pi$ S-wave scattering length parameter $|a_{0}-a_{2}|$.

{\small

\begin{itemize}

\item[{[1]}]
V.~Bernard,  N.~Kaiser and U.~Meissner, Nucl. Phys. B {\bf 357} (1991) 129.

\vsp

\item[{[2]}]
B.~Adeva et al., the DIRAC Collaboration, CERN/SPSC 2000-032, 17 Aug 2000.

\vsp

\item[{[3]}]
S.M.~Bilenky et al., Sov. J. Nucl. Phys. {\bf 10} (1969) 469. 

\end{itemize}
}

\newpage

\setcounter{equation}{0}
\setcounter{figure}{0}\setcounter{table}{0}

\begin{center}
{\Large{\bf Can one extract ${\bf m_{\bf s}/\hat m}$ from 
\mbox{\boldmath$\pi$}K atom properties?}}

\bigskip

{\bf H. Sazdjian}\\[2mm]  

{\em Groupe de Physique Th\'eorique, Institut de Physique Nucl\'eaire,\\  
Universit\'e Paris XI, F-91406 Orsay Cedex, France\\
E-mail: sazdjian@ipno.in2p3.fr}
\end{center}

The lifetime and the $2p-2s$ level energy splitting of the $\pi$K atom
are studied. The lifetime is essentially dependent on the combination
$(a_0^{1/2}-a_0^{3/2})$ of the scattering lengths of the strong
interaction process $\pi K\rightarrow \pi K$. This combination is
independent, at leading order of chiral perturbation theory, of the
quark mass ratio $m_s/\hat m$. The measurement of the $\pi K$ atom lifetime
would therefore mainly provide information about higher-order (one-loop
level) low energy constants. The energy splitting between the $2p-2s$ levels
is essentially dependent on the combination $(2a_0^{1/2}+a_0^{3/2})$ of
the scattering lengths. This combination depends, at leading order of
chiral perturbation theory, on the quark mass ratio $m_s/\hat m$. This
ratio is however multiplied by a term proportional to the deviation of the
$SU(2)\times SU(2)$ quark condensate from the Gell-Mann--Oakes--Renner
value. Therefore, if at the $SU(2)\times SU(2)$ level ($\pi\pi$
scattering amplitude) the chiral perturbation theory is close on
experimetal grounds to the standard scheme of Gell-Mann, Oakes and Renner
and of Gasser and Leutwyler, then the $m_s/\hat m$ dependence of the $2p-2s$
energy splitting would hardly be detectable. On the other hand, if sizable
deviations from the standard scheme are already present at the
$SU(2)\times SU(2)$ level, then these are amplified in the $2p-2s$
energy splitting by the quark mass ratio and experimental detection
of the $m_s/\hat m$ dependent term becomes possible. In any event, the
measurement of the $2p-2s$ energy splitting in the $\pi$K atom would
provide an experimental test for the standard scheme of chiral
perturbation theory. In general, the measurements of the lifetime and
of the energy splitting would allow the determination of each of the
scattering lengths $a_0^{1/2}$ and $a_0^{3/2}$.

\newpage

\setcounter{equation}{0}
\setcounter{figure}{0}\setcounter{table}{0}

\begin{center}
{\Large{\bf \boldmath{$\pi K$} scattering and dispersion relations}}
\bigskip

{\bf B.~Ananthanarayan$^1$, \underline{P.~B\"uttiker$^2$} and 
B.~Moussallam$^3$}\\[2mm]  

$^1$ {\em Centre for Theoretical Studies, Indian Institute of Science, 
Bangalore 560 012,
India}

$^2$ {\em Institut f\"ur Kernphysik, Forschungszentrum J\"ulich, 52425 
J\"uelich,
Germany}

$^3$ {\em Groupe de Physique Th{\'e}orique, IPN, Universit{\'e} Paris--Sud, 
91406 Orsay
C{\'e}dex, France}
\end{center}

Chiral perturbation theory provides the low energy effective theory of
the standard model that describes the interactions involving hadronic
degrees of freedom.
As a non--renormalizable theory, it requires additional low energy
constants (LEC) that have to be introduced at each order of the loop--
or momentum--expansion. In $SU(3)$ at next--to--leading order, ten
LECs $L_i(\mu)$ enter the calculations. 
However, these quantities cannot
be fixed by chiral symmetry only; experimental information is
needed to pin them down. Using the $K_{l4}$ form
factors and $\pi\pi$ sum rules provide accurate estimates for some of the
LECs (e.g. $L_1, L_2, L_3$) but leave the large $N_c$--suppressed LEC
$L_4$  essentially undetermined. This is due to the fact that in such
processes, contributions from $L_4$ accidentally are suppressed.
Such a suppression does not occur in $\pi K$
scattering [1]. Therefore this
process is suitable for a more reliable determination of $L_4$ and also
provides estimates for $L_1, L_2$ and $L_3$. We use dispersion
relations to fix these quantities [2], relying on analyticity,
unitarity and crossing symmetry only, which are the suitable tools for a
comparison of experimental data with ChPT. By combining fixed--$t$
and hyperbolic dispersion relations for $T^+(s,t,u)$ and $T^-(s,t,u)$
and rewriting the chiral amplitudes, we find an appropriate matching of the
chiral and the dispersive representations of these amplitudes.
Saturating the dispersive integrals with the available data [3], we
find the following estimates
\vspace*{-0.5cm}
\begin{center}
\begin{minipage}{5cm}
\begin{eqnarray*}
   L_1^r(m_\rho) & = & 0.84 \pm 0.15,\\
   L_3^r(m_\rho) & = & -3.65 \pm 0.45,
\end{eqnarray*}
\end{minipage}\hspace*{1cm}
\begin{minipage}{5cm}
\begin{eqnarray*}
   L_2^r(m_\rho) & = & 1.36 \pm 0.13,\\
   L_4^r(m_\rho) & = & 0.22 \pm 0.3,
\end{eqnarray*}
\end{minipage}
\end{center}
yielding an estimate for $L_4$ with improved reliability.

\bigskip

{\small

\begin{itemize}

\item[{[1]}]
V.~Bernard, N.~Kaiser and U.--G.~Mei{\ss}ner, Phys.~Rev.~D~{\bf 43},
2757 (1991); Nucl. Phys. B {\bf 357}, 129 (1991).

\vsp

\item[{[2]}]
B.~Ananthanarayan and P.~B\"uttiker, Eur. Phys. J. C {\bf 19}, 517
(2001) [hep-ph/0012023]; B.~Ananthanarayan, P.~B\"uttiker and
B.~Moussallam, Eur. Phys. J. C (2001), to appear.

\vsp

\item[{[3]}]
P.~Estabrooks et al., Nucl.~Phys.~B {\bf 133}, 490 (1978); D.~Aston et
al., Nucl.~Phys.~B {\bf 296}, 493 (1988); D.~Cohen et al., Phys.~Rev.~D
{\bf 22}, 2595 (1988); A.~Etkin et al., Phys. Rev. D
{\bf 25}, 1786 (1982)

\end{itemize}
}

\newpage

\setcounter{equation}{0}
\setcounter{figure}{0}\setcounter{table}{0}

\begin{center}
{\Large{\bf Isospin violation in pion--kaon scattering}}

\bigskip

{\bf \underline{B. Kubis} and U.-G. Mei{\ss}ner}\\[2mm]  

{\em Forschungszentrum J\"ulich, 
Institut f\"ur Kernphysik (Theorie), D--52425 J\"ulich, Germany}

\end{center}

Pion--kaon scattering near threshold is one of the cleanest processes to
test our understanding of chiral dynamics in the presence of strange quarks. 
The existing determinations of the
S--wave scattering lengths are, however, plagued by large uncertainties,
such that high hope lies in the extraction of these parameters
from $\pi K$ bound state experiments [1].

In order to relate the lifetime of $\pi^- K^+$ (or $\pi^+ K^-$) atoms to
one particular combination of the pion--kaon scattering lengths, one has
to make use of a modified Deser formula for the decay into the neutral 
$\pi^0 K^0$
channel which includes next--to--leading order effects in isospin breaking,
\begin{equation}
\Gamma_{\pi^0K^0} \propto \left( a_0^{3/2} - a_0^{1/2} + \epsilon \right)^2 
(1+K) 
\label{deser}
\end{equation}
(see [2] for an equivalent formula for the case of pionium). 
Here, $\epsilon$ represents the isospin violating correction in 
the scattering amplitude $\pi^- K^+ \to \pi^0 K^0$ at threshold, while $K$ 
is an additional contribution only calculable within the bound state 
formalism.

We have evaluated the parameter $\epsilon$ up to one--loop order, including
leading effects both in the quark mass difference $m_u - m_d$ and the 
electromagnetic
fine structure constant $\alpha = e^2/4\pi$ [3] (see also [4]).
While $m_u - m_d$ effects are forbidden at leading order for $\pi\pi$ 
scattering and pionium
and thus numerically negligible in comparison to electromagnetic effects, 
there are
contributions linear in $m_u - m_d$ to the $\pi K$ scattering lengths which 
are of comparable size to the corrections of electromagnetic origin.
At one--loop order, virtual photon loops induce a kinematical divergence at 
threshold (the Coulomb pole) $\propto 1/|\bf{q}_{\rm in}|$ which has to be 
subtracted
in order to properly define the scattering length. Furthermore, photon loops
contain an infrared divergence, 
which however vanishes at threshold. The corresponding radiative
process $\pi^- K^+ \to \pi^0 K^0 \gamma$ which has to be included in order 
to define an infrared finite total cross section yields a totally negligible
contribution at threshold.  

The combined result for the scattering length, expressed in terms of relative
corrections 
to the tree level isospin symmetric one, numerically reads 
\begin{eqnarray}
a_0(\pi^- K^+ \to \pi^0 K^0) &=& 
\frac{\sqrt{2}}{3} \left(a_0^{3/2}-a_0^{1/2} \right)^{\!\rm tree}
 \biggl\{\,1 
         \,+\! \underbrace{0.006}_{{\cal O}(m_u-m_d)}      
         \!+\,\, \underbrace{0.009 }_{{\cal O}(e^2)} 
\nonumber\\ && \quad
           +\, \underbrace{(0.121 \pm 0.009)}_{{\cal O}(p^4)} 
         \,+\, \underbrace{(0.008 \pm 0.002)}_{{\cal O}(p^2(m_u-m_d))}    
         \,-\, \underbrace{(0.009 \pm 0.008)}_{{\cal O}(p^2 e^2)} \, 
\biggr\} ~. 
\end{eqnarray}
The error ranges displayed for the next--to--leading order contributions are
 due 
 to 
the uncertainties in the low--energy constants entering at that order. 
The value for the lifetime of the $\pi K$ atom obtained using the
 ``uncorrected''
Deser formula, $\tau_{\pi K} = (4.8 \pm 0.1) \times 10^{-15}$~s, 
is slightly reduced to $\tau_{\pi K} = (4.7 \pm 0.2) \times 10^{-15}$~s.
The additional correction $K$ in (\ref{deser}) however remains to be
 calculated.

{\small

\begin{itemize}

\item[{[1]}]
B. Adeva et al., CERN-SPSC-2000-032, CERN-SPSC-P-284-ADD-1, Aug 2000.

\vsp

\item[{[2]}]
A.~Gall et al., 
Phys.\ Lett.\ B \textbf{462} (1999) 335; 
J.~Gasser et al.,
Phys.\ Lett.\ B \textbf{471} (1999) 244.

\vsp

\item[{[3]}]
B.~Kubis and U.-G.~Mei{\ss}ner, arXiv:hep-ph/0107199, Nucl.\ Phys.\ A, 
in print.

\vsp

\item[{[4]}]
A.~Nehme and P.~Talavera, arXiv:hep-ph/0107299. 

\end{itemize}
}

\newpage

\setcounter{equation}{0}
\setcounter{figure}{0}\setcounter{table}{0}

\begin{center}
{\Large{\bf The first measurement of kaonic nitrogen 

with DEAR at DA$\Phi$NE}}

\bigskip

{\bf C. Curceanu (Petrascu)$^{1}$}\\

{on behalf of the DEAR Collaboration}\\[2mm]

$^1$ {\em  LNF-INFN, Via E. Fermi 40, 00044 Frascati (Roma), Italy 

({\small on leave from IFIN-HH, Bucharest, Romania})}

\end{center}

In May 2001 the DEAR (DA$\Phi$NE Exotic Atom Research) collaboration [1]
performed the first measurement of an exotic atom (kaonic nitrogen) at the
DA$\Phi$NE collider of the Laboratori Nazionali di Frascati dell'INFN.
The measurement was performed with a cryogenic target (118 K) filled
with nitrogen at about 2 bar ($\rho \simeq 4.5 \rho_{NTP}$). 
As detector,  8 CCD-22 were used.
A total integrated luminosity of about 2 pb$^{-1}$ was collected.
After an optimization of the machine performances and of the DEAR degrader,
an X-ray spectrum was obtained and analysed by performing a global fit.
The global fit (with a $\chi^2$/NDOF = 0.95) gave a statistical
significance of the two kaonic nitrogen lines measurable with DEAR of 
about 3$\sigma$ for the line at 4.58 keV ($7\to 6$ transition) and 
of about 3.7$\sigma$ for the
line at 7.6 keV ($6\to 5$ transition), 
giving a global statistical significance of about
5$\sigma$. These results compare favourable with the Monte Carlo
simulation results with an yield of the two transitions of about 
90$^{+10}_{-25}\%$.
An empty target-like measurement was as well performed.
The background subtracted spectrum is presented in Figure 1.
The experiment could therefore demonstrate the capability of creating and
detecting exotic atoms using the kaons flux from DA$\Phi$NE.

Next stage of the experiment is the measurement of the K-complex in 
kaonic hydrogen.

\begin{figure}[thb]
\begin{center}
\mbox{\epsfig{file=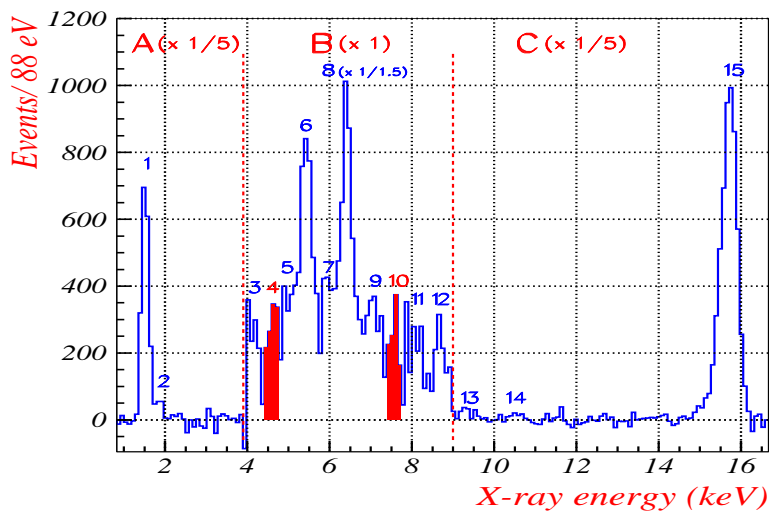,height=7.7cm,width=12cm}}
\caption{\small{The kaonic nitrogen background subtracted spectrum. X-ray
    lines are: 1 = Al K$_{\alpha}$; 2 = Si K$_{\alpha}$; 3 = Sc K$_{\alpha}$;
4 = {\bf KN (4.58 keV)}; 5 = V K$_{\alpha}$; 6 = Cr  K$_{\alpha}$;
7 = Cr K$_{\beta}$ + Mn K$_{\alpha}$; 8 = Fe K$_{\alpha}$ + Mn K$_{\beta}$;
9 = Fe K$_{\beta}$; 10 = {\bf KN (7.62 keV)}; 11 = Cu K$_{\alpha}$;
12 = Zn K$_{\alpha}$; 13 = Zn K$_{\beta}$; 14 = Pb L$_{\alpha}$  }}
\end{center}
\end{figure}

\vspace*{-0.3cm}

{\small
\begin{itemize}
\item[{[1]}]
S.~Bianco {\it et al.}, Rivista del Nuovo Cimento {\bf 22},
No. 11 (1999) 1.
\end{itemize}
}

\newpage


\begin{center}
{\Large{\bf The quark condensate on the lattice}}

\bigskip

{\bf P. Hasenfratz}\\[2mm]
 
{\em Institut f\"ur Theoretische Physik, Universit\"at Bern, 
Sidlerstr. 5, 3012 Bern, Switzerland} 

\end{center}

Chiral symmetry in QCD with $N_f \ge 2$ massless quark flavours is
expected to be broken spontaneously producing a non-zero $\langle{\bar
\psi}\psi\rangle$ quark condensate generated through
non-perturbative effects. A direct study of this condensate requires
chiral symmetric lattice fermions (domain-wall, overlap, or the
fixed-point formulations) which are expensive to simulate. The fermion
condensate
\begin{equation}
\langle{\bar \psi}\psi\rangle
=-\lim_{m_q \rightarrow 0}\lim_{V \rightarrow \infty} \frac{d}{dm_q}\log Z
\end{equation}
is related to the low energy constant $\Sigma$ of the tree level
effective chiral lagrangean as $\langle{\bar \psi}\psi\rangle = -N_f \Sigma$.
A way to determine the low energy constant $\Sigma$ is to measure the
condensate in a small volume V in a fixed topological sector $Q$ of QCD:
$N_f\Sigma_{m_q,V,Q}= -\langle{\bar \psi}\psi\rangle_{m_q,V,Q}$ and
use chiral perturbation theory to connect it to $\Sigma$ defined at
$m_q=0$ and $V=\infty$. Until now this is carried through in the
quenched approximation only. Chiral perturbation theory predicts in
this case~[1]
\begin{equation}
\lim_{m_q \rightarrow 0}\frac{\Sigma_{m_q,V,Q}-|Q|/m_qV}{m_q}=
\frac{\Sigma^2 V}{2|Q|} + \frac{c}{a^2}\,,
\label{3}
\end{equation}
where $|Q|/m_qV$ is the contribution from the exact zero modes, while
$c/a^2$ ($a$ is the lattice unit) is the ultraviolet divergent cut-off
effect induced by the finite quark mass. Measuring $\Sigma_{m_q,V,Q}$
in different volumes, eq.~(\ref{3}) allows to find $\Sigma$. The bare
low energy constant $\Sigma$ is multiplicatively renormalized. The
renormalization factor between $\Sigma$ and $\Sigma_{\overline {MS}}(\mu)$ is
calculated by combining non-perturbative measurements and continuum
perturbation theory. We obtained the result
$\Sigma_{\overline {MS}}(2 {\rm GeV})=(270 \pm 10 {\rm MeV})^3$~[2],
which, using the GMOR relation, corresponds to ${\overline
m}_{\overline {MS}}(2 {\rm GeV})= (3.9 \pm 0.4) {\rm MeV}$ for the averaged 
u,d
quark mass. This result is obtained at a resolution $a
\approx 0.13 {\rm fm}$ and no systematic continuum extrapolation is
performed yet. The same applies also to the other measurements (they are
consistent with our result) available at this moment~[3]. One should
keep in mind that these results refer to the quenched approximation
($N_f=0$). It is unknown how far is the $N_f=0$ low energy
constant from that at $N_f=3$ . The number above (which is close to the
phenomenologically 
expected one) suggests that $\Sigma(N_f)$ depends smoothly on
$N_f$. This is not true, however for the condensate $\langle{\bar
\psi}\psi\rangle$ itself. The $N_f=0$ (quenched) QCD is not a healthy
QFT and the tree level relation 
$\langle{\bar \psi}\psi\rangle = -N_f \Sigma$, unlike in full QCD,
receives corrections which are even diverging in the chiral
limit. As opposed to the low energy constant $\Sigma$, the condensate
is not well defined in the quenched approximation~[1].

\bigskip

{\small

\begin{itemize}

\item[{[1]}]
J.C.~Osborn et al., Nucl. Phys. B540 (1999) 317; P.M.~Damgaard et
al., Nucl. Phys. B547 (1999) 305.

\vsp

\item[{[2]}]
P.~Hasenfratz, S.~Hauswirth, K.~Holland, T.~J\"org and F.~Niedermayer,
arXiv:hep-lat/0109007. 

\vsp

\item[{[3]}]
P.~Hernandez et al., JHEP 0107 (2001) 018; T.~DeGrand,
arXiv:hep-lat/0107014; L.~Giusti et al., arXiv:hep-lat/0108007.

\end{itemize}

}

\newpage

\setcounter{equation}{0}
\setcounter{figure}{0}\setcounter{table}{0}

  \begin{center}
  {\Large{\bf Chiral and lattice calculations of the $\pi \pi$ scattering
  lengths}} 

  \bigskip 
 
  {\bf Gilberto Colangelo}\\[2mm]   
  {\em Institut f\"ur Theoretische Physik
der Universit\"at Z\"urich
Winterthurerstr. 190 CH-8057 Z\"urich}
  \end{center}

\bigskip

In this talk I have reviewed the status of lattice calculations of the pion
scattering lengths from a nonexpert point of view. The history of such
calculations starts in the eighties/early nineties. I have split the
discussion by reviewing first the calculations pre-2000 and then those
post-2000. The most advanced calculation of last century is that of
Fukugita et al. [1], where reference to earlier literature can also be
found. This paper is also particularly relevant because, as far as I know,
it is the only attempt at calculating the scattering length in the $I=0$
channel on the lattice. Despite the huge effort and the remarkable results,
various systematic effects were not under good control.

A new generation of calculations by two japanese collaborations have
started in recent years [2,3]. Although the final
results are not yet available, encouraging and very interesting preliminary
results for the $I=2$ scattering lengths have been presented at recent
Lattice conferences. 

I have also discussed the recent results of a small chinese collaboration
that has exploited the use of improved actions to do the calculation on a
very coarse and anisotropic lattice [4], and therefore perform it on a
relatively small computer. I sincerely hope that this brave and encouraging
calculation will challenge also smaller lattice groups to attempt such
nontrivial but extremely interesting calculation.

The agreement between experiments, CHPT, and lattice calculations for the
$\pi \pi$ scattering lengths would represent a tremendous achievement for
our understanding of strong interactions
The CHPT calculation has reached a very high level of accuracy [5],
and recent experimental results have confirmed the prediction [6].
Hopefully, the lattice calculations will also reach the same level of
accuracy one day.

\bigskip

{\small

\begin{itemize}

\item[{[1]}]
M.~Fukugita, Y.~Kuramashi, M.~Okawa, H.~Mino and A.~Ukawa,
Phys.\ Rev.\ D {\bf 52} (1995) 3003 [arXiv:hep-lat/9501024].

\vsp

\item[{[2]}]
S.~Aoki {\it et al.}  [JLQCD Collaboration],
Nucl.\ Phys.\ Proc.\ Suppl.\  {\bf 83} (2000) 241\\~
[arXiv:hep-lat/9911025].

\vsp

\item[{[3]}]
S.~Aoki {\it et al.}  [CP-PACS Collaboration],
arXiv:hep-lat/0110151.

\vsp

\item[{[4]}]
C.~Liu, J.h.~Zhang, Y.~Chen and J.P.~Ma, arXiv:hep-lat/0109020;
arXiv:hep-lat/0109010.

\vsp

\item[{[5]}]
G.~Colangelo, J.~Gasser and H.~Leutwyler,
Phys.\ Lett.\ B {\bf 488} (2000) 261 [arXiv:hep-ph/0007112].

\vsp

\item[{[6]}]
S.~Pislak {\it et al.}  [BNL-E865 Collaboration],
Phys.\ Rev.\ Lett.\  {\bf 87} (2001) 221801 \\~
[arXiv:hep-ex/0106071].

\end{itemize}
}

\newpage

\setcounter{equation}{0}
\setcounter{figure}{0}\setcounter{table}{0}

\begin{center}
{\Large{\bf The quark condensate from $K_{e4}$ decays}}

\bigskip

{\bf G.~Colangelo$^1$,
{\underline{J.~Gasser}$^2$} and  
H.~Leutwyler$^2$}\\[2mm]

{\begin{tabular}{c}
$^1\,$Institute for Theoretical Physics, University of 
Z\"urich\\
Winterthurerstr. 190, CH-8057 Z\"urich, Switzerland\\
$^2\,$Institute for Theoretical Physics, University of 
Bern\\   
Sidlerstr. 5, CH-3012 Bern, Switzerland
\end{tabular}  }

\end{center}

Roy equations [1] combined with chiral symmetry allow one to predict
[2] the
low-energy behaviour of the $\pi\pi$ scattering amplitude with high
precision. The prediction is based on the hypothesis that the quark
condensate is the leading order parameter of spontaneously broken
chiral symmetry. This has been questioned  in Refs. [3,4], where it is
emphasized that experimental evidence for this scenario is not
available. The recent result from the high statistics $K_{e4}$
experiment E865 at Brookhaven [5] closes this gap, because those data
allow one to determine the size of the leading term in the quark mass
expansion of the pion mass. E865 confirms [5] the hypothesis that
underlies our prediction of the  $\pi\pi$ $S$-wave scattering lengths:
 more than 94\% of the pion mass stems from the quark condensate. The
generalized framework of $SU(2)\times SU(2)$ chiral perturbation theory
developed in [3], that allows for a small or vanishing condensate,
 can therefore be dismissed.

\vskip.2cm

{\em Note added}: In Ref. [6], the E865 data [5] have been 
analyzed without the use of chiral symmetry constraints.

\bigskip

{\small

\begin{itemize}
\item[{[1]}] 
S.M.~Roy, Phys.\ Lett.\ B {\bf 36} (1971) 353.

\vsp

\item[{[2]}]
G.~Colangelo, J.~Gasser and H.~Leutwyler,
Phys.\ Lett.\  B {\bf 488} (2000) 261
[hep-ph/0007112]; 
 Nucl. Phys. B {\bf 603} (2001) 125 [hep-ph/0103088];
Phys.\ Rev.\ Lett.\  {\bf 86}, 5008 (2001)
[hep-ph/0103063].

\vsp

\item[{[3]}]
M.~Knecht, B.~Moussallam, J.~Stern and N.H.~Fuchs,
Nucl.\ Phys.\ B {\bf 457} (1995) 513 
[hep-ph/9507319];
{\it ibid.} B {\bf 471} (1996) 445 
[hep-ph/9512404].

\vsp

\item[{[4]}]
L.~Girlanda, M.~Knecht, B.~Moussallam and J.~Stern,
Phys.\ Lett.\  B {\bf 409} (1997) 461 
[hep-ph/9703448].

\vsp

\item[{[5]}]
S.~Pislak {\it et al.}  [BNL-E865 Collaboration],
Phys.\ Rev.\ Lett.\  {\bf 87} (2001) 221801 
[hep-ex/0106071].

\vsp
\item[{[6]}]
S.~Descotes, N.H.~Fuchs, L.~Girlanda and J.~Stern,
arXiv:hep-ph/0112088.

\end{itemize}

}

\newpage

\setcounter{equation}{0}
\setcounter{figure}{0}\setcounter{table}{0}

\begin{center}
{\Large{\bf Radiative corrections to semileptonic decays}}

\bigskip

{\bf M. Knecht}\\[2mm]  

{\em Centre de Physique Th\'eorique, CNRS Luminy Case 907, 
F-13288 Marseille Cedex 9}

\end{center}

Semileptonic decays of the light pseudoscalar mesons  provide important
informations on the non-perturbative aspects of the QCD vacuum
(low-energy $\pi\pi$ scattering and low-energy constants $L_i$ 
from $K_{\ell 4}$ decay), 
or on the CKM matrix elements $V_{ud}$ (pion $\beta$ decay) and $V_{us}$
($K_{\ell 3}$ decay). An accurate determination of these quantities requires
high statistics experiments, as well as reliable theoretical descriptions.
In particular, isospin breaking effects, arising both from $m_u\neq m_d$ and 
from radiative corrections, should be included. So far, this has not been 
done in a systematic way.

A general framework allowing for a systematic treatment of isospin breaking 
effects within the chiral low-energy expansion has been introduced in [1], 
and applied to the simplest, $\pi_{\ell 2}$ and $K_{\ell 2}$ semileptonic 
modes. Recently, the $K_{\ell 3}$ channels have been treated as well [2] 
(for a preliminary discussion of the $K_{\ell 4}$ modes, see [3]). 

In the presence of radiative corrections, the factorization between the 
hadronic and leptonic currents in semileptonic decays is lost. In the case
of the $K_{\ell 3}$ transitions, this entails that the two form factors 
$f_{\pm}$ not only depend on the momentum transfer between the two mesons, 
but also on a second, independent, kinematical variable. Furthermore, the 
necessity to include the decay mode with a soft photon emission, in order 
to obtain an infrared finite result, modifies the structure of the Dalitz 
plot density. Explicit expressions for the $K_{\ell 3}$ form factors 
at one loop with isospin breaking effects included have been obtained in [2],
where existing expressions [4] for the Dalitz plot distribution have 
been corrected for several mistakes.

The effects of isospin breaking on the extraction of $V_{us}$ along the 
lines described in [5] have also been discussed and found to be small 
(see [2] for details). A reliable estimate of the two-loop strong 
interaction effects appears more than ever as being the crucial issue for a 
precise determination of $V_{us}$ from $K_{\ell 3}$ decay [6].

\bigskip

{\small

\begin{itemize}

\item[{[1]}]
M. Knecht,  H. Neufeld, H. Rupertsberger and  P. Talavera,
Eur.\ Phys.\ J.\ C {\bf 12} (2000) 469 [hep-ph/9909284].

\vsp

\item[{[2]}]
V. Cirigliano, M. Knecht, H. Neufeld, H. Rupertsberger and P. Talavera,
submitted to Eur.\ Phys.\ J.\ C [hep-ph/0110153].

\vsp

\item[{[3]}]
V. Cuplov, contribution to this meeting.

\vsp

\item[{[4]}]
E.S.~Ginsberg, Phys.\ Rev.\ {\bf 142} (1966) 1035; 
{\bf 162} (1967) 1570
[Erratum-ibid.\  {\bf 187} (1967) 2280];
 {\bf 171} (1968) 1675
[Erratum-ibid.\  {\bf 174} (1968) 2169]; 
Phys.\ Rev.\ D {\bf 1} (1970) 229.

\vsp

\item[{[5]}]
H. Leutwyler and M. Roos, Z.\ Phys.\ C {\bf 25} (1984) 91.

\vsp

\item[{[6]}]
N.H.~Fuchs, M.~Knecht and J.~Stern,
Phys.\ Rev.\ D {\bf 62} (2000) 033003 [hep-ph/0001188]. 

\end{itemize}
}

\newpage


\setcounter{equation}{0}
\setcounter{figure}{0}\setcounter{table}{0}

\begin{center}

{\Large{\bf Isospin violation in $K_{\ell 4}$ decays}}

\bigskip

{\bf Vesna Cuplov}\\[2mm]  

{\em Centre de Physique Th\'eorique, CNRS Luminy, Case 907,
 F-13288 Marseille Cedex 9, France} 

\end{center}

The framework used in Ref.[1] for studying $K_{\ell 4}$ decays is
 modified in order to account for isospin violating contributions,
 $m_u \neq m_d$ and $\alpha \neq 0$.
The full kinematics, which requires three reference 
frames and five variables introduced by Cabibbo and Maksymowicz
 Ref.[2] is changed. It is necessary to introduce a tensorial form 
factor, $T_{\mu \nu} = \frac{1}{M_K^2}Tp_{1\mu} p_{2\nu}$, in the 
matrix element:
\begin{eqnarray}
{\mathcal T} &=& \frac{G_F V^\star_{us}}{\sqrt{2}}~{\biggl[}~\bar{u} 
(p_\nu) \gamma_\mu(1-\gamma_5) v(p_l) (V^\mu - A^\mu) +~ \bar{u} (p_\nu) 
\sigma_{\mu \nu} (1+\gamma_5) v(p_l) T^{\mu \nu} ~{\biggr]} \nonumber
\end{eqnarray}
However this tensorial form factor appears only in the $\pi^+ \pi^-$ 
channel, due to electromagnetic corrections. \\
The decay rate is $d \Gamma_5 = G^2_F \mid V_{us} \mid^2  N(s_\pi, s_l) J_5 
(s_\pi, s_l, \theta_\pi, \theta_l, \phi) \times ds_\pi ds_l 
d (\cos \theta_\pi) d(\cos \theta_l) d\phi $. We have a general formula for 
$J_5$ written in terms of the mass of the kaon which decays, the masses of 
the two pions in the final state and all the possible form factors, 
including the tensorial one. We explicitly calculate the form factors for
 $K_{\ell 4}$ at tree level using an effective Lagrangian, where not only 
the pseudoscalars but also the photon and the light leptons have been 
included as dynamical degrees of freedom Ref.[3].\\
The numerical values obtained for the decay rates(MeV) and phase 
space($\textrm{MeV}^3$) are shown in the table below. We compare to the
 values obtained by Bijnens and al. Ref.[4].

\vspace*{.2cm}

\renewcommand{\arraystretch}{0.70}

{\small
\noindent\begin{tabular}{c | c c | c c | c c}
{\footnotesize Decay} 
&\hspace*{-0.1cm} {\footnotesize (1) $\pi^+\pi^-e^+\nu_e$}\hspace*{-0.1cm} 
&\hspace*{-0.1cm} {\footnotesize $\pi^+\pi^-\mu^+\nu_\mu$}\hspace*{-0.1cm} 
&\hspace*{-0.1cm} {\footnotesize (2) $\pi^o\pi^oe^+\nu_e$}\hspace*{-0.1cm} 
&\hspace*{-0.1cm} {\footnotesize $\pi^o\pi^o\mu^+\nu_\mu$}\hspace*{-0.1cm} 
&\hspace*{-0.1cm} {\footnotesize (3) $\pi^o\pi^-e^+\nu_e$}\hspace*{-0.1cm} 
&\hspace*{-0.1cm} {\footnotesize $\pi^o\pi^-\mu^+\nu_\mu$}\hspace*{-0.1cm} \\
\hline
{\footnotesize Ref.[4]} &&&&&\\
{\footnotesize Tree level} & 1297 & 155 & 683 & 102 & 561 & 55 \\
\hline
\hspace*{-0.25cm}{\footnotesize Masses as in Ref.[4]}\hspace*{-0.25cm} 
&&&&&\\
$m_u=m_d$ & 1292 & 154 & 681 & 101 & 561 & 55 \\
$e=0~T=0$ &&&&&\\
{\footnotesize phase space MeV$^3$} & 1.140 & 0.146 & 0.674 & 0.099 & 1.339 
& 0.197 \\
\hline
{\footnotesize Physical masses} &&&&&\\  
$m_u\neq m_d$ & 1292 & 153 & 730 & 110 & 543 & 51 \\
$e\neq 0~T\neq 0$ &&&&&\\
{\footnotesize phase space MeV$^3$} & 1.140 & 0.146 & 0.674 & 0.099 & 1.307 
& 0.187 \\
\end{tabular} 
}

\vspace*{.4cm}

\noindent
In Ref.[4], the masses of the particles are $(M_\pi,M_K)$=(139.6, 493.6) MeV,
(135, 493.6) MeV and (137, 497.7) MeV for the decays (1), (2) and (3),
respectively.\\ 
In case of no isospin violation, our Fortran routine reproduces the results 
of
Bijnens and al. at better than 0.5 $\%$. 
The interpretation of these numerical values leads to the result that the
isospin violation, at tree level, is due to form factors for the channel
$\pi^o \pi^o$ and to phase space for the channel $\pi^o \pi^-$. The channel
$\pi^+ \pi^-$ is the only one which involves the tensorial form factor. At
tree level, the effects of isospin violation and the existence of a tensorial
form factor are numerically tiny in this channel. The calculations at one 
loop
level are in progress. 

\bigskip

{\small

\begin{itemize}

\item[{[1]}]
The second Da$\Phi$ne physics handbook - Vol.1.

\vsp

\item[{[2]}]
N. Cabibbo and A. Maksymovicz, Phys. Rev. {\bf 137} (1965) B438; erratum 
Phys. Rev. {\bf 168} (1968) 1926.

\vsp

\item[{[3]}]
M. Knecht, H. Neufeld, H. Rupertsberger and P. Talavera, 
Eur. Phys. J. C {\bf 12} (2000) 469.

\vsp

\item[{[4]}]
J. Bijnens, G. Colangelo and J. Gasser, 
Nucl. Phys. B {\bf 427} (1994) 427.  
 
\end{itemize}
}

\newpage

\setcounter{equation}{0}
\setcounter{figure}{0}\setcounter{table}{0}

\begin{center}
{\Large{\bf Resonance estimates of low-energy
constants in chiral Lagrangians and QCD short-distance constraints}}

\bigskip

{\bf M. Knecht and \underline{A. Nyf\/feler}}\\[2mm]  

{\em Centre de Physique Th\'{e}orique, CNRS-Luminy, Case 907, F-13288
Marseille Cedex 9, France} 
\end{center}

At higher orders in the chiral expansion and when radiative
corrections are included, a large number of unknown low-energy
constants enters in ChPT. Lacking enough data, one often relies on
estimates for these constants using some resonance Lagrangian. This
approach was quite successful for the $L_i$ at order $p^4$ [1],
although one has to make sure that QCD short-distance constraints are
met [2]. Starting from the study of the low-energy and high-energy
behaviors of the QCD three-point functions $\langle V\!AP\rangle$,
$\langle VVP\rangle$ and $\langle AAP\rangle$, we have evaluated in
[3] several ${\cal O}({p}^6)$ low-energy constants of the chiral
Lagrangian within the framework of the lowest meson dominance (LMD)
approximation to the large-$N_C$ limit of QCD. In certain cases,
values that differ substantially from estimates based on a resonance
Lagrangian are obtained:
\begin{center}
\begin{minipage}[t]{8.7cm}
\renewcommand{\arraystretch}{1.1}
\begin{tabular}{|l|r@{.}l|r@{.}l|r@{.}l|r@{.}l|r@{.}l|r@{.}l|}
\hline
 & \multicolumn{2}{|c|}{{$C_{78}$}}  
 & \multicolumn{2}{|c|}{{$C_{82}$}}  
 & \multicolumn{2}{|c|}{{$C_{87}$}}  
 & \multicolumn{2}{|c|}{{$C_{88}$}}  
 & \multicolumn{2}{|c|}{{$C_{89}$}}  
 & \multicolumn{2}{|c|}{{$C_{90}$}}  
\\ 
\hline  
LMD    & 1 & 09 & -0 & 36 & 0 & 40 & -0 & 52 & 1 & 97 
& 0 & 0  \\ 
Set I  & 1 & 09 & -0 & 29 & 0 & 47 & -0 & 16 & 2 & 29
& 0 & 33 \\
Set II & 1 & 49 & -0 & 39 & 0 & 65 & -0 & 14 & 3 & 22
&  0 & 51 \\
\hline
\end{tabular}
\end{minipage}\hspace*{-0.7cm} 
\parbox[t]{5mm}{\rule[0mm]{0mm}{0mm}} 
\raisebox{3.6mm}{\begin{minipage}[t]{8.3cm}
\begin{tabular}{|l|r@{.}l|r@{.}l|r@{.}l|r@{.}l|}
\hline  
 & \multicolumn{2}{|c|}{{$A_{V,p^2}$}}  
 & \multicolumn{2}{|c|}{{$A_{V,(p+q)^2}$}}  
 & \multicolumn{2}{|c|}{{$A_{A,p^2}$}}  
 & \multicolumn{2}{|c|}{{$A_{A,(p+q)^2}$}}  
\\ \hline  
LMD   & -1 & 11 &\hspace*{2mm} -0 & 26 & -0 & 14 &\hspace*{2mm} 0 & 040 \\
Set I & -1 & 13 &   0 & 0 & -0 & 096 &  0 &  0 \\
\hline
\end{tabular}
\rule[-1cm]{0mm}{1cm} 
\end{minipage}\hspace*{-0.3cm} }
\end{center} 
In the table we have collected our estimates (LMD) for the low-energy
constants $C_i$ (even parity) from [4] and $A_i$ (odd parity) from
[5], all in units of $10^{-4} / F_0^2$, and compared them to those
obtained with a resonance Lagrangian, using different sets of
parameters in the Lagrangian (set I and II).  We have introduced the
abbreviations $A_{V,p^2} = A_2 - 4 A_3$, $A_{V,(p+q)^2} = A_2 - 2 A_3
- 4 A_4$, $A_{A,p^2} = A_{11} + A_{23} - 4 A_{24}$, and $A_{A,(p+q)^2}
= 3 A_{11} + 4 A_{17} + A_{23} - 2 A_{24} - 4 A_{25}$.

The differences, in particular in $C_{88}, C_{90}, A_{V,(p+q)^2},$ and
$A_{A,(p+q)^2}$, arise through the fact that QCD short-distance
constraints are in general not correctly taken into account in the
approaches using resonance Lagrangians. Whereas in the case of the
vector form factor of the pion our result for a certain combination of
low-energy constants at order $p^6$ agrees with the one [6] from the
resonance Lagrangian, our estimate for one of the counterterm
contributions in the form factor $A$ in the decay $\pi\to e \nu_e
\gamma$ is by a factor of 5 smaller than the resonance estimate in [7]. 

Finally, in Ref. [3] we also briefly discussed how to systematically
improve our approach. For instance, for the pion-photon-photon
transition form factor it is necessary to include a second vector
resonance (LMD+V) in order to correctly reproduce the CLEO data [8]
for the form factor at high $Q^2$, with one photon on-shell. In this
way one obtains the new estimate $A_{V,p^2}^{LMD+V} = - 1.36 \times
10^{-4}/F_0^2$, i.e., about 20\% away from the LMD estimate given in
the table.

\bigskip

{\small

\begin{itemize}

\item[{[1]}]
G.~Ecker, J.~Gasser, A.~Pich and E.~de Rafael,
Nucl.\ Phys.\ B {\bf 321} (1989) 311.

\vsp

\item[{[2]}]
G.~Ecker, J.~Gasser, H.~Leutwyler, A.~Pich and E.~de Rafael,
Phys.\ Lett.\ B {\bf 223} (1989) 425.

\vsp

\item[{[3]}]
M.~Knecht and A.~Nyf\/feler,
Eur.\ Phys.\ J.\ C {\bf 21} (2001) 659. 

\vsp

\item[{[4]}]
J.~Bijnens, G.~Colangelo and G.~Ecker,
JHEP {\bf 9902} (1999) 020; [hep-ph/9902437].

\vsp

\item[{[5]}]
H.W.~Fearing and S.~Scherer, 
Phys.\ Rev.\ D {\bf 53} (1996) 315. 

\vsp

\item[{[6]}]
J.~Bijnens, G.~Colangelo and P.~Talavera,
JHEP {\bf 9805} (1998) 014; [hep-ph/9805389].

\vsp

\item[{[7]}]
J.~Bijnens and P.~Talavera,
Nucl.\ Phys.\ B {\bf 489} (1997) 387. 

\vsp

\item[{[8]}]
J.~Gronberg {\it et al.}  [CLEO Collaboration],
Phys.\ Rev.\ D {\bf 57} (1998) 33. 

\end{itemize}
}

\newpage

\setcounter{equation}{0}
\setcounter{figure}{0}\setcounter{table}{0}

\begin{center}
{\Large{\bf The new pionic hydrogen experiment at PSI}}

\bigskip

{\bf M.\,Hennebach} for the $\pi$H collaboration (www.pihydrogen.psi.ch) 
\\[2mm]  

{\em Institut f\"ur Kernphysik, Forschungszentrum J\"ulich, D-52425 J\"ulich}
\end{center}

A new high--precision determination of the strong--interaction shift
($\epsilon_{1s}$) and broadening ($\Gamma_{1s}$) of the ground state in 
pionic
hydrogen ($\pi \mathrm{H}$) has been started at PSI [1]. This constitutes a
direct measurement of the $\pi \mathrm{N}$ scattering lengths a$^+$ and a$^-$
and is an important test of the methods of chiral perturbation theory. 

The experimental approach is to determine the energies of x-rays emitted by
radiative transitions to the 1s--state in pionic hydrogen. This state is both
shifted in energy because of the strong interaction between proton and pion
and broadened because its lifetime is shortened due to the additional decay
channels. The energy of the {\it np--1s} transitions in $\pi$H is in the 
range
of a few keV and should be determined with an accuracy in the 20 meV
range. This can be achieved only with a crystal spectrometer. 

A previous experiment at PSI [2] measured the {\it 3p--1s} transition in
pionic hydrogen and deuterium ($\pi$D), combining a highly efficient stopping
arrangement (cyclotron trap) for the pion beam and a cylindrically bent Bragg
crystal with newly developed CCD (Charge Coupled Device) detectors. While an
accuracy of 0.5\% was reached for $\epsilon_{1s}$, the error for 
$\Gamma_{1s}$
(8\%) was an order of magnitude higher, which makes it hard to determine both
a$^+$ and a$^-$ from the $\pi$H measurement alone. The $\pi$D shift
measurement can be used to improve the situation, but this necessitates a 
good
theoretical understanding of the 3--body--problem in the deuterium system. 

The goal of this experiment is to reduce the uncertainty of $\Gamma_{1s}$ to
$\approx$1\% and that of $\epsilon_{1s}$ to 0.2\%. This can be achieved by
using an improved cyclotron trap, spherically bent crystals and a new
large--area CCD for a higher x-ray rate and better background suppression. 

During the first production run the {\it 3p-1s} transition was measured at
three different pressures (4\,bar, 28\,bar and liquid hydrogen) to 
investigate
the possible influence of mechanisms that depend on the collision rate 
between
$\pi$H atoms and target molecules. One process is the formation of molecular
states ($\mathrm{\pi^-H + H_2 \to [(\pi^-pp)p]_{ee}}$) that could skew the
$\epsilon_{1s}$ measurement. The other is Coulomb deexcitation
($\mathrm{\pi^-H_{(n)} + H_2 \to \pi^-H_{(n')} + H_2}$), a non--radiative
process that leads to a Doppler broadening of the line. 

A preliminary analysis of the data (Table\,1) does not show a pressure
dependence of the shift $\epsilon_{1s}$. Furthermore, the averaged result 
does
not deviate from the earlier result [2] which gave $\epsilon_{1s}$ =
7108\,$\pm$\,13\,$\pm$\,34 at 15\,bar. The width $\Gamma_{1s}$ (including the
Coulomb deexcitation) appears to be pressure dependent. This will be
investigated in the next stage of the experiment, starting in Spring 2002. 

\vspace{-3mm}
\begin{table}[ht]
\small
\begin{center}
\begin{tabular}{|c|c|c|}
\hline
\raisebox{0mm}[4mm][2mm]{Target pressure} & $\epsilon_{1s}$ [meV] &
$\Gamma_{1s}$ [meV]\\ 
\hline\hline
\multicolumn{1}{|l|}{\raisebox{0mm}[4mm]{4\,bar}} & 
7082\,$\pm$\,31\,$\pm$\,15
& 973\,$\pm$\,75\,$\pm$\,10\\[0.5mm] 
\hline
\multicolumn{1}{|l|}{\raisebox{0mm}[4mm]{28\,bar}} &
7137\,$\pm$\,18\,$\pm$\,40 & 969\,$\pm$\,26\,$\pm$\,10\\[0.5mm] 
\hline
\multicolumn{1}{|l|}{\raisebox{0mm}[4mm]{liquid ($\sim$700\,bar)}} &
7095\,$\pm$\,25\,$\pm$\,25 & 1052\,$\pm$\,58\,$\pm$\,10\\[0.5mm] 
\hline
\end{tabular}
\vspace{-1mm}
\caption{\small Preliminary fit results for the Summer 2001 data
  (value\,$\pm$\,stat.\,$\pm$\,sys.)} 
\vspace{2mm}
\end{center}
\end{table}

\vspace{-10mm}

{\small

\begin{itemize}

\item[{[1]}]
Proposal R-98-01.1 at PSI.

\vsp

\item[{[2]}]
H.-Ch.\,Schr\"oder et al., Eur. Phys. J. {\bf C 21} (2001) 433 vol.\,3

\end{itemize}
}

\newpage

\setcounter{equation}{0}
\setcounter{figure}{0}\setcounter{table}{0}

  \begin{center}
  {\Large{\bf Atomic cascade in hadronic atoms}}

  \bigskip

  {\bf\underline{T.S. Jensen}$^{1,2}$ and V.E. Markushin$^{1}$}\\[2mm]  

  $^1$ {\em Paul Scherrer Institute, CH-5232, Villigen PSI, Switzerland}

  $^2$ {\em Institut f{\"u}r Theoretische Physik der Universit{\"a}t
  Z{\"u}rich, 

 Winterthurerstrasse 190,\\ CH-8057 Z{\"u}rich, Switzerland}

  \end{center}
   The atomic cascades in exotic hydrogen atoms $x^-p$ 
($x^-=\mu^-$, $\pi^-$, $K^-$, $\bar{p}$) have been studied 
using a new Monte Carlo code that calculates
the evolution of the $x^-p$ kinetic energy distribution 
from the very beginning of the cascade. The cascade model is based
on new cross sections for the collisional processes: Stark mixing,
elastic scattering, nuclear absorption during collisions, 
Coulomb deexcitation, and external Auger effect.  
For the final part of the cascade ($n\leq 5$), 
Stark mixing, deceleration, and nuclear absorption 
during collisions  for $x^-p$ scattering from
hydrogen atoms have been computed in the close--coupling approximation~[1-3];
this fully quantum mechanical framework takes threshold and strong absorption
effects into account.    
For $n\geq 8$, a classical--trajectory Monte Carlo model [3] for exotic atoms
scattering from molecular hydrogen  has been used to calculate the
cross sections for Stark mixing, elastic scattering, and
Coulomb deexcitation.  The semiclassical (eikonal) approximation is
used for values of $n$ between the quantum mechanical and  classical 
domains.

\begin{figure}[h!]
\parbox{.35\textwidth}{\centerline{\epsfig{file=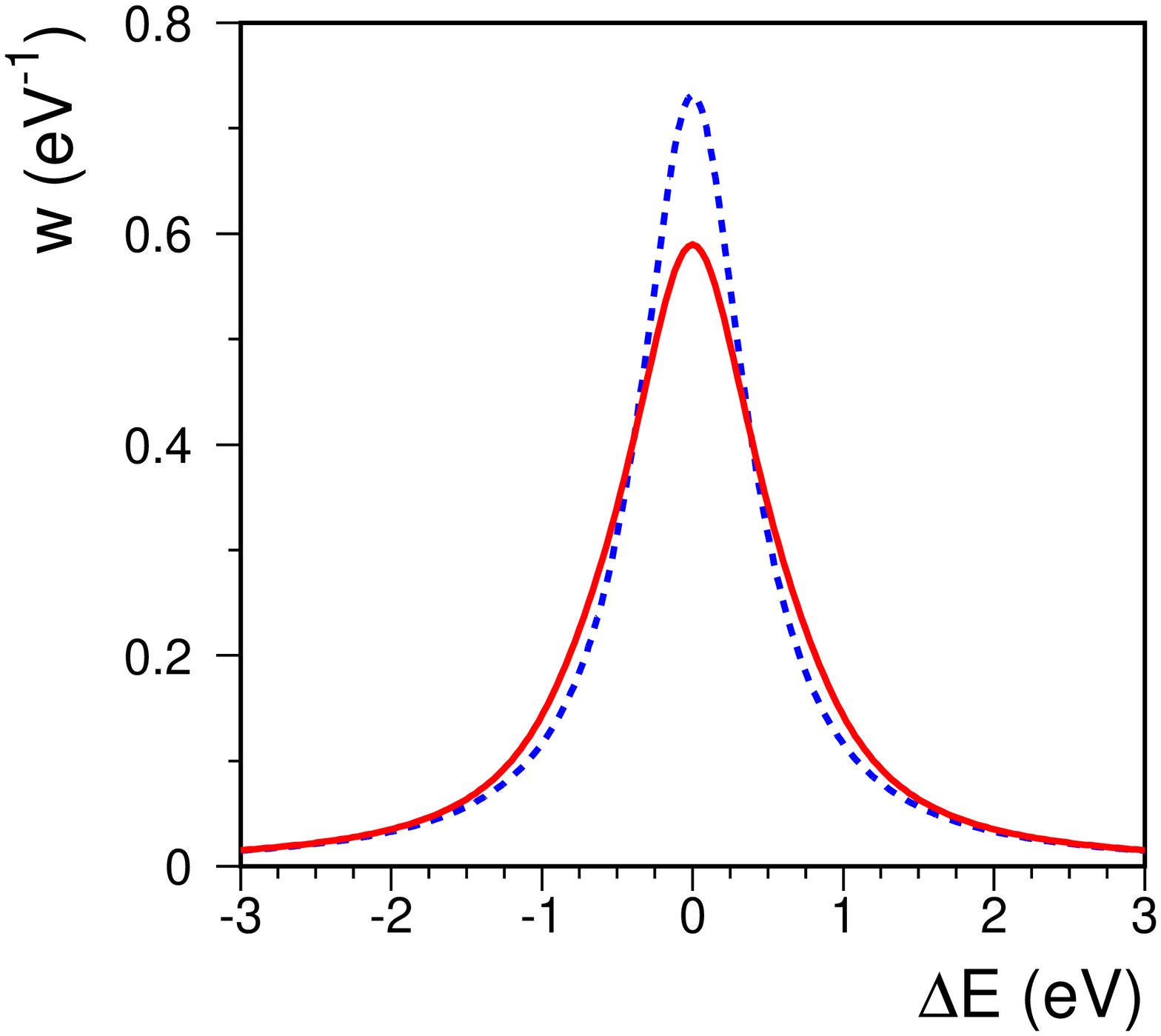,
width=.33\textwidth,silent=,clip=}}}
\hfill \parbox{.60\textwidth}{\caption{\label{jensen-fig}
The calculated energy profile of the $\mathrm{K}_\beta$
line at 3~bar in pionic hydrogen with Doppler broadening taken into account 
(solid line)
in comparison with the natural line shape assuming 
$\Gamma_{1S}=0.87\;$eV (dashed line).
 }}
\end{figure}

The predictions of the detailed kinetics cascade model have  been confronted
with the experimental observables: X-ray yields, cascade times, and 
kinetic energy distributions~[4]. The case of pionic hydrogen is of
particular interest. The new
PSI experiment~[5], whose goal is to determine the strong 
interaction width $\Gamma_{1S}$ of pionic hydrogen on the level of 1\%,
relies crucially on a good understanding of the kinetic
energy evolution during the atomic cascade because 
the measured $nP\to 1S$ line profiles must be corrected for the Doppler
broadening (see Fig.~\ref{jensen-fig}).

\bigskip

{\small

\begin{itemize}

\item[{[1]}]
   T.S.~Jensen and V.E.~Markushin, PSI-PR-99-32 (1999), 
arXiv:nucl-th/0001009.

\vsp

\item[{[2]}]
   T.S.~Jensen and V.E.~Markushin, Nucl. Phys. A {\bf 689} (2001) 537.

\vsp

\item[{[3]}]
   T.S.~Jensen and V.E.~Markushin, Hyperfine Interactions, in press.

\vsp

\item[{[4]}]
   V.E.~Markushin and  T.S.~Jensen, Hyperfine Interactions, in press.

\vsp

\item[{[5]}]
   D.~Gotta,  $\pi N$ Newsletter {\bf 15} (1999) 276. See also 
M. Hennebach, these proceedings.

\end{itemize}
}

\newpage

\setcounter{equation}{0}
\setcounter{figure}{0}\setcounter{table}{0}

\begin{center}
{\Large{\bf Pion-nucleon analysis}}

\bigskip

{\bf M.E. Sainio}\\[2mm]

{\em Helsinki Institute of Physics, and Department of Physical Sciences,
University of Helsinki, P.O. Box 64,
FIN-00014 Helsinki, Finland}
\end{center}

There is information of the pion-nucleon interaction in different charge
channels available at low energy from a variety of
experiments: 1) cross section information, 2) analyzing powers,
3) forward $D^+$ amplitude, 4) pionic hydrogen level shift and width
measurements. Over the years such information has been analysed with
different techniques, e.g.
\begin{enumerate}
\item Karlsruhe fixed-$t$ analysis KH78/80 [1], partial wave dispersion
      relation analysis KA.84 [2] and partial wave relation analysis KA.85 
[3].
\item VPI-GWU analysis with fixed-$t$ constraints, accessible through the
      SAID facility \\{\tt http://gwdac.phys.gwu.edu/}
\item Gashi et al. analysis at very low energies with an effective range
      expansion [4].
\item Low-energy analysis exploiting forward dispersion relations [5].
\end{enumerate}

A major problem in extracting hadronic quantities from observables
is the question of the treatment of the electromagnetic effects. In
low-energy scattering the formalism of Tromborg et al. [6] is often used, but
there are effects not taken into account by that method. Also,
for pionic hydrogen, determining the hadronic scattering lengths requires
a careful analysis of electromagnetic corrections [7,8]. For the
low-energy scattering region a ${\cal O}(p^3)$ calculation in chiral
perturbation theory including all electromagnetic effects to that
order has recently been published [9].

It turns out, especially in the analysis of sensitive quantities like
the $\Sigma$-term, that in addition to the electromagnetic corrections
also the high partial waves are important. Therefore, it is well
motivated to start a new analysis in the spirit of Karlsruhe [10].

\bigskip

{\small

\begin{itemize}

\item[{[1]}]
R.~Koch and E.~Pietarinen, Nucl. Phys. A {\bf 336} (1980) 331.

\vsp

\item[{[2]}]
R.~Koch, Z. Phys. C {\bf 29} (1985) 597.

\vsp

\item[{[3]}]
R.~Koch, Nucl. Phys. A {\bf 448} (1986) 707.

\vsp

\item[{[4]}]
A.~Gashi et al., arXiv:hep-ph/0009081.

\vsp

\item[{[5]}]
J.~Gasser et al., Phys. Lett. B {\bf 213} (1988) 85.

\vsp

\item[{[6]}]
B.~Tromborg at al., Phys. Rev. D {\bf 15} (1977) 725.

\vsp

\item[{[7]}]
D.~Sigg et al., Nucl. Phys. A {\bf 609} (1996) 310.

\vsp

\item[{[8]}]
V.E.~Lyubovitskij and A.~Rusetsky, Phys. Lett. B {\bf 494} (2000) 9.

\vsp

\item[{[9]}]
N.~Fettes and U.-G.~Mei\ss ner, Nucl. Phys. A {\bf 693} (2001) 693.

\vsp

\item[{[10]}]
P.~Piirola et al., arXiv:hep-ph/0110044.

\end{itemize}
}

\newpage

\setcounter{equation}{0}
\setcounter{figure}{0}\setcounter{table}{0}

\begin{center}
{\Large{\bf Theory of low-energy pion-nucleon scattering}}

\bigskip

{\bf Nadia Fettes and \underline{Ulf-G. Mei{\ss}ner}}\\[2mm]  

{\em Forschungszentrum J\"ulich, Institut f\"ur Kernphysik (Theorie),
D-52425 J\"ulich, Germany}

\end{center}
In the Standard Model there are two sources of isospin violation.
Strong isospin violation is driven by the quark mass difference
$(m_u-m_d)$ whereas electromagnetic (em) isospin violation is due to the
different quark charges. To extract the strong contribution, which
leads to small effects of the size $(m_u-m_d)/\Lambda_{\rm QCD} \sim
2\%$, one thus has to be able to simultaneously treat both sources.
In the two flavor sector of QCD, pion-nucleon scattering is best
suited to study isospin violation since the isovector operator 
$(m_u-m_d)(\bar u u - \bar d d)$ contributes at the same order as
the isoscalar chiral symmetry breaking one, $(m_u+m_d)(\bar u u +
\bar d d)$. In the isospin limit, the $\pi$N scattering amplitude
decomposes into isoscalar and isovector parts. This allows to
formulate  a variety of triangle relations between  measurable
channels that vanish if isospin were an exact symmetry, see e.g.[1].
As a first necessary step, one has to consider the isospin limit.
In the framework of heavy baryon CHPT, the S- and P-wave phases
can be described precisely in a fourth order calculation [2]. Fixing the
low-energy constants (LECs) for pion lab momenta 
between 50 and 100~MeV by fitting to the
existing $\pi$N partial wave analyses (KH, VPI/GWU, Z\"urich), one can 
predict
threshold parameters and the phases at higher energies. Most
importantly this calculation shows convergence of the chiral expansion.
The inclusion of virtual photons (denoted as $\gamma^\ast$)
in the effective $\pi$N Lagrangian is
straightforward, for a detailed discussion see e.g.~[3]. In [4], a
complete third order calculation was performed including one- and
two-photon exchange diagrams, photon loop corrections, soft photon 
radiation and three em LECs, which parameterize the effects
of hard photons and vacuum polarization. A fit to all data (with the
exception of the ones generally considered inconsistent) below 
pion lab momenta of 100~MeV allowed to determine the strong and em
LECs and to uniquely separate the hadronic and the em amplitudes. A very
pronounced difference in the hadronic amplitude for
elastic scattering $\pi^- p \to \pi^- p$
compared to the standard PWA's was observed. This difference could be
traced back to the inclusion of non-linear $\pi \pi \bar N N \gamma^\ast$
couplings as demanded by chiral symmetry. Such effects have to be
included and thus a revision of the commonly used em corrections like
e.g. the NORDITA approach is called for. In the S--wave triangle relation
between elastic scattering, $\pi^\pm p \to \pi^\pm p$, and
charge exchange, $\pi^- p \to \pi^0 n$, one finds  strong isospin
violation of $-0.75\%$ in the low-energy region,
an order of magnitude smaller than reported in the literature [5] but
consistent we the expected size of isospin violation. It is now of
utmost importance to map these result obtained in chiral perturbation
theory on a dispersive representation to analyze also $\pi$N data at
higher energies.

\bigskip

{\small

\begin{itemize}

\item[{[1]}]
N.~Fettes, Ulf-G.~Mei{\ss}ner and S.~Steininger, Phys. 
Lett. B \textbf{451} (1999) 233;
N.~Fettes and Ulf-G.~Mei{\ss}ner, Phys.Rev. C \textbf{63} (2001) 045201.

\vsp

\item[{[2]}]
N.~Fettes and  Ulf-G.~Mei{\ss}ner, Nucl. Phys. A \textbf{676} (2000) 311.

\vsp

\item[{[3]}]
G. M\"uller and Ulf-G. Mei{\ss}ner, Nucl. Phys. B {\bf 556} (1999) 265.

\vsp

\item[{[4]}]
N.~Fettes and  Ulf-G.~Mei{\ss}ner,  Nucl. Phys. A \textbf{693} (2001) 693.

\vsp

\item[{[5]}]
W.R.~Gibbs, Li~Ai, and W.B.~Kaufmann,   Phys. Rev. Lett. 
{\bf 74} (1995) 3740; E. Matsinos,  Phys. Rev. C {\bf 56} (1997) 3014.

\end{itemize}
}

\end{document}